\documentclass[useAMS,usenatbib]{mn2e}

\usepackage{graphicx}
\usepackage{natbib}

\newcommand{\ltsima} {$\; \buildrel < \over \sim \;$}
\newcommand{\gtsima} {$\; \buildrel > \over \sim \;$}
\newcommand{\lta} {\lower.5ex\hbox{\ltsima}}
\newcommand{\gta} {\lower.5ex\hbox{\gtsima}}

\title{On the optimality of the spherical Mexican hat wavelet estimator for the primordial non-Gaussianity}

\author[A. Curto et al.]{A. Curto,$^1$\thanks{e-mail:
curto@ifca.unican.es} E. Mart\'{\i}nez-Gonz\'alez,$^1$ R. B. Barreiro$^1$\\
$^1$     Instituto de F\'isica de Cantabria, CSIC-Universidad de Cantabria, Avda. de los Castros s/n, 39005 Santander, Spain.\\
}
\date{Accepted  Received ; in original form }

\begin{document}

\maketitle

\begin{abstract}
We study the spherical Mexican hat wavelet (SMHW) as a detector of
primordial non-Gaussianity of the local type on the Cosmic Microwave
Background (CMB) anisotropies. For this purpose we define third order
statistics based on the wavelet coefficient maps and the original
map. We find the dependence of these statistics in terms of the
non-linear coupling parameter $f_{nl}$ and the bispectrum of this type
of non-Gaussianity. We compare the analytical values for these
statistics with the results obtained with non-Gaussian simulations for
an ideal full-sky CMB experiment without noise.  We study the power of
this method to detect $f_{nl}$, i. e. the variance of this parameter
$\sigma^2(f_{nl})$, and compare it with the variance obtained from the
primary bispectrum for the same experiment. Finally we apply our
wavelet based estimator on WMAP-like maps with incomplete sky and
inhomogeneous noise and compare with the optimal bispectrum estimator.
The results show that the wavelet cubic statistics are as efficient as
the bispectrum as optimal detectors of this type of primordial
non-Gaussianity.
\end{abstract}
\section{Introduction}
The primordial perturbations generated during the inflationary period
are imprinted in the radiation and matter distribution. The study of
the CMB anisotropies has become an important source of information to
understand the physics of the very early universe. Thus for example
the search of primordial non-Gaussianities on the CMB anisotropies,
the measurement of the tilt and running of the index of the power
spectrum $\Delta^2_{\mathcal{R}}(k) = A k^{n_s-1}$ and the search of
primordial gravitational waves have become part of a set of
observables that are being used to select among many different models
for the inflationary paradigm. The simplest models of inflation as the
standard, single-field, slow roll inflation
\citep{guth,albrecht,linde1982,linde1983,mukhanov1992} predict that
the anisotropies are compatible with a nearly Gaussian random field.

There are two main procedures for the Gaussianity analyses of a map of
CMB anisotropies. One can perform blind tests searching for deviations
with respect to the null hypothesis (the random field is
Gaussian). The second option is to consider different physical
scenarios and to look for their imprints on the anisotropies.

Among the many different blind tests performed on the {\it Wilkinson
  Microwave Anisotropy Probe} WMAP\footnote{http://map.gsfc.nasa.gov/}
data, there are several reports of anomalies present in these data. We
can mention the asymmetry between the two ecliptic hemispheres
\citep{eriksen2004,hansen2004,eriksen2005,eriksen2007,hofttuft2009,pietrobon2010,vielva2010},
anomalous quadrupole-octupole alignment
\citep{copi2004,oliveira2004,copi2006,gruppuso2009,frommert2010}, a
non-Gaussian cold spot
\citep{vielva2004,mukherjee2004,cruz2005,cruz2006,cruz2007,cruz2007b,cruz2008,vielva2010a},
unexpected alignment of CMB structures \citep{wiaux2006,vielva2007}
and an unexpected low value of the CMB variance
\citep{monteserin2008,cruz2010}.

Regarding targeted tests, one has to think about the possible physical
mechanisms that lead to non-Gaussianities and search for their
possible signatures. Historically motivated, many inflationary models
that generate non-Gaussianity can be parametrised by the local
non-linear coupling parameter $f_{nl}$, which is introduced through
the primordial gravitational potential
\citep{verde2000,komatsu2001,bartolo2004}
\begin{equation}
\Phi({\bf x}) = \Phi_L({\bf x}) + f_{nl}\{\Phi_L^2({\bf x})-\langle \Phi_L({\bf x}) \rangle^2 \}.  
\label{phi_vs_fnl}
\end{equation}
Significant non-Gaussianity of the local form can be generated for
example in the curvaton model \citep{lyth2003}, multi-field
inflationary models \citep{komatsu2005}, the inhomogeneous reheating
scenario \citep{dvali2004,bartolo2004}, models with low reheating
efficiency \citep{mukhanov1992,salopek1990}, models based on hybrid
inflation \citep{lin2009}, etc. This kind of non-Gaussianity is
characterised by correlations among modes $k$ in the Fourier space
with very different amplitude
\citep{creminelli2006,creminelli2007}. This can be seen for example
through the shape function $F(k_1,k_2,k_3)$ of the local
non-Gaussianity
\begin{equation}
F(k_1,k_2,k_3) = A f_{nl}  \left(\frac{1}{k_{1}^3k_{2}^3} + \frac{1}{k_{1}^3k_{3}^3} + \frac{1}{k_{2}^3k_{3}^3}\right)
\end{equation}
where $A$ is a normalization constant \citep[see for example the plot
  of the shape function of the local distribution
  by][]{babich2004}. The most significant contributions of this kind
of non-Gaussianity arise for the cases with $k_1 << k_2\approx k_3$
and permutations among the three modes $k_1$, $k_2$ and $k_3$. Other
shapes of inflationary models that produce their particular kind of
non-Gaussianity are for example the {\it equilateral} and the {\it
  orthogonal} shape \citep{senatore2010,komatsu2010}. The most
significant contribution to the non-Gaussianity is located in specific
ranges of the Fourier space $k_1\approx k_2\approx k_3$ for the {\it
  equilateral} shape whereas the {\it orthogonal} shape is nearly
orthogonal to the two previous forms.

In this paper we focus on the non-Gaussianity of local type. The
canonical inflationary model predicts $f_{nl} \sim 10^{-2}$ whereas
other models predict larger amounts of non-Gaussianity
\citep{bartolo2004}. An eventual detection of a deviation from
Gaussianity will rule out many inflationary models from the present
available list. Many studies have been performed to constrain the
local $f_{nl}$ on different data sets such as the {\it Cosmic
  Background Explorer (COBE)} data \citep{komatsu2002,cayon2003}, the
MAXIMA data \citep{santos2003,cayon2003b}, the {\it Very Small Array}
data \citep{smith2004}, the BOOMERang data \citep{troia,natoli2009}
and the Archeops data \citep{curto2007,curto2008}. Significant
improvements on the $(S/N)$ ratio of the local $f_{nl}$ parameter have
been achieved with the WMAP data using different estimators. For most
of the works, $f_{nl}$ is positive with a significance between
$1 \sigma$ and $2 \sigma$. See for example the results
obtained with different bispectrum-based estimators
\citep{komatsu2003,spergel2007,creminelli2006,creminelli2007,wandelt2008,komatsu2009,smith2009,elsner2009,bucher2009,komatsu2010,smidt2010},
with the SMHW \citep{curto2009a,curto2009b}, with needlets
\citep{pietrobon2009,rudjord2009,rudjord2010,cabella2010}, with the
HEALPix wavelet \citep{casaponsa2010}, with the Minkowski functionals
\citep{hikage2006,gott2007,hikage2008,matsubara2010}, the N-PDF
distribution \citep{vielva2009}, the skewness of the power spectrum
\citep{smidt2009}, etc. Other works use the distribution of matter on
large scales \citep[see for
  example][]{dalal2008,matarrese2008,slosar2008,seljak2009,desjacques2010,xia2010}
to constrain the local $f_{nl}$.

This work is a study of the efficiency of the third order
wavelet-based estimators \citep{curto2009a,curto2009b} to detect
primordial non-Gaussianity of the local type. We find the dependence
of the third order estimators in terms of the local bispectrum and
$f_{nl}$. We compare the power of this method to detect $f_{nl}$ with
the optimal bispectrum estimator for an ideal full-sky and noiseless
experiment and the same comparison for an experiment with WMAP 5-year
and WMAP 7-year beam, noise and sky cut properties. Our results
indicate that the wavelet estimators are as efficient as the optimal
estimators based on the bispectrum.

The article is organised as follows. Section
\ref{the_third_order_stat} presents the estimators based on wavelets,
their dependence on the angular bispectrum, their analytical
covariance matrix and the $f_{nl}$ Fisher matrix for wavelet and
bispectrum estimators. In Section \ref{comparison_with_opt_bisp_est}
we compare the $\sigma(f_{nl})$ values obtained with the wavelet
estimator and the bispectrum estimator and the conclusions are
presented in Section \ref{conclusions}.
\section{The third order statistics}
\label{the_third_order_stat}
\subsection{Expected values of the wavelet estimator in terms of the bispectrum}
The third order statistics of this analysis are based on the SMHW. See
\citet{antoine1998,martinez2002,vielva2007b,martinez2008} for detailed
information about the wavelets and a list of applications to the CMB
anisotropies. Given a function $f(\bf n)$ defined at a position $\bf
n$ on the sphere and a continuous wavelet family on that space
$\Psi({\bf n}; {\bf b}, R)$, we define the continuous wavelet
transform as
\begin{equation}
w(R; {\bf b}) = \int d{\bf n}f({\bf n})\Psi({\bf n}; {\bf b}, R)
\label{wavmap}
\end{equation}
where ${\bf b}$ is the position on the sky at which the wavelet
coefficient is evaluated and $R$ is the scale of the wavelet.

\begin{figure}
  \center
  \includegraphics[height=5.0cm,width=8.0cm]{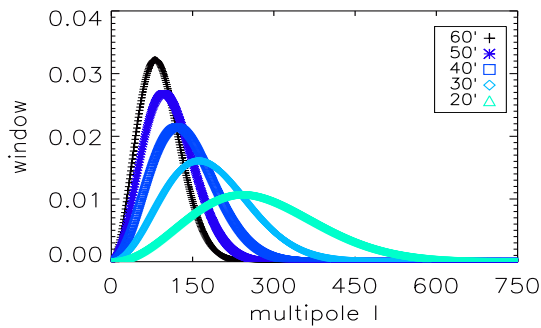}
  \caption{The window function for the SMHW at different angular
    scales. Note that the wavelet filters high multipoles
    for low angular scales and viceversa.}
  \label{waveletwindow}
\end{figure}

Considering a set of different angular scales $\{R_i\}$ we define a
third order statistic depending on three scales $\{i,j,k\}$
\citep{curto2009b}
\begin{equation}
q_{i j k}=\frac{1}{4\pi}\frac{1}{\sigma_i\sigma_j\sigma_k}\int d{\bf n} w(R_i,{\bf n})w(R_j,{\bf n})w(R_k,{\bf n})
\label{themoments_qijk}
\end{equation}
where $\sigma_i$ is the dispersion of the wavelet coefficient map
$w(R_i,{\bf n})$. In the particular case of $R_0=0$, $w(R_0,{\bf n})
\equiv f({\bf n})$. Using the properties of the wavelet, we have
\begin{equation}
w(R_i,{\bf n})=\sum_{\ell m}a_{\ell m}\omega_\ell (R_i)Y_{\ell m}(\bf n)
\label{thewaveletconv}
\end{equation}
and
\begin{equation}
\sigma_i^2=\sum_{\ell}C_{\ell}\frac{2\ell+1}{4\pi}\omega_\ell^2(R_i)
\label{thewaveletdisp}
\end{equation}
where $\omega_\ell (R)$ is the window function of the wavelet at a
scale $R$ and it is given by the harmonic transform of the mother
wavelet of the SMHW \citep{martinez2002,sanz2006}.  The convolution
with the wavelet is equivalent to filter the maps with a window
function $\omega_l(R)$ which depends on the scale. In
Fig. \ref{waveletwindow} we plot the wavelet window function for
several angular scales. For small scales, the wavelet filters low
multipoles and viceversa. Therefore it is important to select a set of
angular scales that ranges all the interesting multipoles.
\subsection{The statistics and the primordial non-Gaussianity}
\label{statistics_and_fnl}
Considering Eqs. \ref{themoments_qijk} and \ref{thewaveletconv}, the
third order moments can be written as
\begin{eqnarray}
\nonumber
q_{i j k}=\frac{1}{4\pi}\frac{1}{\sigma_i\sigma_j\sigma_k} \\
\nonumber
\times \Bigg\{ \sum_{\ell_1,\ell_2,\ell_3,m_1,m_2,m_3}a_{\ell_1m_1}a_{\ell_2m_2}a_{\ell_3m_3} \\
\nonumber
\times \omega_{\ell_1}(R_i)\omega_{\ell_2}(R_j)\omega_{\ell_3}(R_k)  \\ 
\nonumber
\times \left ( \begin{array}{ccc} \ell_1 & \ell_2 & \ell_3 \\ m_1 & m_2 & m_3 \end{array}\right ) \left ( \begin{array}{ccc} \ell_1 & \ell_2 & \ell_3 \\ 0 & 0 & 0 \end{array}\right ) \\
\times \sqrt{\frac{(2\ell_1+1)(2\ell_2+1)(2\ell_3+1)}{4\pi}} \Bigg\},
\label{main}
\end{eqnarray}
where matrix is the Wigner 3j symbol and we have used the Gaunt
integral \citep{komatsu2001}
\begin{eqnarray}
\nonumber
\int d^2 {\bf n} Y_{\ell_1 m_1}({\bf n})Y_{\ell_2 m_2}({\bf
  n})Y_{\ell_3 m_3}({\bf n}) = \\ 
\nonumber
\times \left ( \begin{array}{ccc} \ell_1 &
  \ell_2 & \ell_3 \\ m_1 & m_2 & m_3 \end{array}\right ) \left
( \begin{array}{ccc} \ell_1 & \ell_2 & \ell_3 \\ 0 & 0 &
  0 \end{array}\right ) \\
\times \sqrt{\frac{(2\ell_1+1)(2\ell_2+1)(2\ell_3+1)}{4\pi}}.
\end{eqnarray}
The mean value of the third order statistic $q_{ijk}$ can be written
in terms of the reduced bispectrum as defined by \citet{komatsu2001}:
\begin{eqnarray}
\nonumber
\langle q_{i j k} \rangle=\frac{1}{4\pi}\frac{1}{\sigma_i\sigma_j\sigma_k} \\
\times \sum_{\ell_1,\ell_2,\ell_3}\omega_{\ell_1}(R_i)\omega_{\ell_2}(R_j)\omega_{\ell_3}(R_k)I_{\ell_1\ell_2\ell_3}^2b_{\ell_1\ell_2\ell_3},
\end{eqnarray}
where $I_{\ell_1\ell_2\ell_3}$ is defined as
\begin{eqnarray}
\nonumber
I_{\ell_1\ell_2\ell_3} = \\
\left ( \begin{array}{ccc} \ell_1 & \ell_2 &
  \ell_3 \\ 0 & 0 & 0 \end{array}\right )
\sqrt{\frac{(2\ell_1+1)(2\ell_2+1)(2\ell_3+1)}{4\pi}}.
\end{eqnarray}
Now assuming a primordial gravitational potential of the form given by
Eq. \ref{phi_vs_fnl} it is possible to derive its corresponding
bispectrum in terms of $f_{nl}$ \citep[see][]{komatsu2001}. The
expected value of the third order moments is proportional to $f_{nl}$
\begin{equation}
\langle q_{i j k} \rangle_{f_{nl}} = \alpha_{i j k} \times f_{nl},
\label{qijk_vs_fnl}
\end{equation}
where 
\begin{eqnarray}
\nonumber
\alpha_{i j k} =
\frac{1}{4\pi}\frac{1}{\sigma_i\sigma_j\sigma_k} \\
\times \Big\{ \sum_{\ell_1,\ell_2,\ell_3}\omega_{\ell_1}(R_i)\omega_{\ell_2}(R_j)\omega_{\ell_3}(R_k)I_{\ell_1\ell_2\ell_3}^2
b^{prim}_{\ell_1\ell_2\ell_3}\Big\}.
\label{thealpha_stats}
\end{eqnarray}
The pixel properties are taken into account by replacing $C_{\ell}$
by $C_{\ell} \left[ \omega_{\ell}^{(pix)} \right]^2$ in Eq.
\ref{thewaveletdisp} and $b_{\ell_1\ell_2\ell_3}^{prim}$ by
$b_{\ell_1\ell_2\ell_3}^{prim} \omega_{\ell_1}^{(pix)}
\omega_{\ell_2}^{(pix)} \omega_{\ell_3}^{(pix)}$ in
Eq. \ref{thealpha_stats}, where $\omega_{\ell}^{(pix)}$ is the pixel
window function for the HEALPix pixelization \citep{healpix}.

\begin{figure}
  \center
  \includegraphics[height=5.0cm,width=8.0cm]{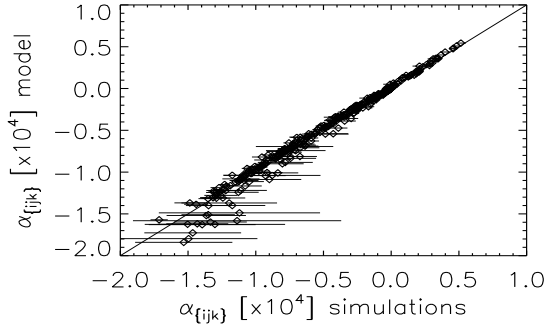}
  \caption[The $\alpha_{i j k}$ statistics]{The $\alpha_{i j k}$
    statistics computed analytically and with $n_{sim}=300$
    non-Gaussian simulations. The error bars correspond to the
    dispersion $\sigma(\alpha_{i j k})/\sqrt{n_{sim}}$ obtained with
    the simulations. The $\alpha_{i j k}$ are sorted such as the
    largest values correspond to pixel-dominated scales while the
    smallest values correspond to the largest scales.}
  \label{fig_thealpha_stats}
\end{figure}

We have evaluated the $\alpha_{i j k}$ statistics analytically using
Eq. \ref{thealpha_stats} as well as from non-Gaussian simulations for
the same set of 12 angular scales used in
\citet{curto2009b}\footnote{The angular scales used in
  \citet{curto2009b} are: 6.9, 10.6, 16.3, 24.9, 38.3, 58.7, 90.1,
  138.3, 212.3, 325.8 and 500 arcmin. The unconvolved map was also
  included in the analysis.}. We have used a full-sky ideal experiment
without noise and a characteristic angular resolution of $6.9$ arcmin
(HEALPix $N_{side}=512$). We need the expected values of the local
primordial bispectrum to evaluate analytically the statistics
$\alpha_{i j k}$. We have computed the primordial bispectrum up to
$\ell_{max}=1535$ using the
gTfast\footnote{http://gyudon.as.utexas.edu/$\sim$komatsu/CRL/nongaussianity/}
code based on CMBFast \citep{seljak1996} to evaluate the transfer
function. The cosmological parameters for this analysis are
$\Omega_{cdm}=0.25$, $\Omega_{b}=0.05$, $\Omega_{\Lambda}=0.70$, $\tau
= 0.09$, $h=0.73$ and a scale invariant spectral index $n=1$ for the
power spectrum $P(k)$. We have used a set of 300 non-Gaussian
simulations generated with the same cosmological parameters following
the algorithm defined in \citet{liguori2003,liguori2007}. The mean
value of the $\alpha_{i j k}$ statistics of these simulations and its
error bars are plotted in Fig. \ref{fig_thealpha_stats}. These values
are compared with the theoretical value obtained from Eq.
\ref{thealpha_stats}. We can see that there is a good agreement
between the simulations and the analytic model. There is a slight
discrepancy at negative values but we do not consider it very
significant due to the correlations among the $\alpha_{ijk}$
quantities. In any case, these values correspond to $\alpha_{ijk}$
involving large scales where cosmic variance is more important and a
larger number of simulations would be needed to achieve
convergence. There is also a slight deviation at small angular scales
(corresponding to values of $\alpha_{ijk} \sim 0.5 \times
10^{-4}$). This may be related to numerical uncertainties in the
evaluation of the integrals that lead to the bispectrum
$b^{prim}_{\ell_1\ell_2\ell_3}$ at high multipoles $\ell$. However, we
have seen that these small deviations do not introduce significant
differences ($< 1\%$) in the estimation of $\sigma(f_{nl})$ when we
use either the analytic or the simulated $\alpha_{ijk}$.
\subsection{The covariance of the third order statistics}
\begin{figure}
  \center
  \includegraphics[height=5.0cm,width=8.0cm]{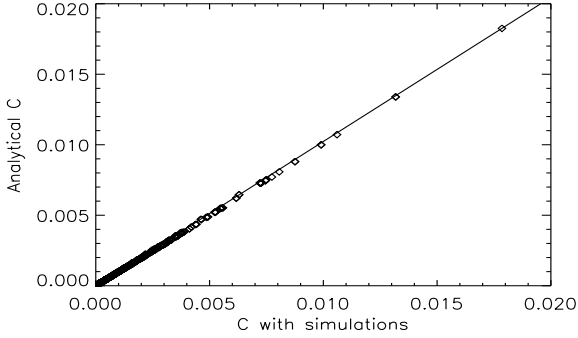}
  \caption{The analytical covariance matrix computed through
    Eq. \ref{the_cov_qijk_stat} compared to the one obtained from
    Gaussian simulations for the same power spectrum $C_{\ell}$.
\label{the_analytical_sims_cov_matrix}}
\end{figure}
The covariance matrix of the third order moments can be computed in
the Gaussian limit using the properties of the covariance for the
bispectrum.  Considering the Gaussian case we have
\begin{equation}
C_{ijk,rst} = \langle q_{ijk} q_{rst}\rangle - \langle q_{ijk} \rangle \langle q_{rst}\rangle = \langle q_{ijk} q_{rst}\rangle.
\end{equation}
From the definition of $q_{ijk}$
\begin{eqnarray}
\nonumber 
\langle q_{ijk} q_{rst}\rangle =
\frac{1}{(4\pi)^2}\frac{1}{\sigma_i\sigma_j\sigma_k}\frac{1}{\sigma_r\sigma_s\sigma_t} \times \int  {\bf d\hat n_1 d\hat n_2}   \\
\nonumber
\times \big \langle w(R_i,{\bf
n_1})w(R_j,{\bf n_1})w(R_k,{\bf n_1}) \\
\times  w(R_r,{\bf n_2})w(R_s,{\bf n_2})w(R_t,{\bf n_2}) \big \rangle.
\label{covq}
\end{eqnarray}
Using Wick's theorem and the properties of Gaussian distributions
\citep[in a similar manner as in Eq. 13 of][]{heavens1998} we have
\begin{eqnarray}
\nonumber
\big \langle w(R_i,{\bf n_1})w(R_j,{\bf n_1})w(R_k,{\bf n_1}) \\
\nonumber
\times w(R_r,{\bf
n_2})w(R_s,{\bf n_2})w(R_t,{\bf n_2})\big \rangle = \\
\nonumber
 \langle w(R_i,{\bf n_1})w(R_j,{\bf n_1})\rangle \\
\nonumber
\times \langle w(R_k,{\bf n_1}) w(R_r,{\bf n_2})\rangle \\
\nonumber
\times \langle w(R_s,{\bf n_2})w(R_t,{\bf n_2})\rangle + \\
 +~permutations~(total~15~terms)
\end{eqnarray}
From these terms, there are only 6 that do not vanish in Eq.
\ref{covq}. They are those which involve the two coordinates ${\bf
  n_1}$ and ${\bf n_2}$ in the same average
\citep{srednicki1993}. Taking this into account and the properties of
the two point correlation functions \citep[see Eq. 14
  of][]{heavens1998} we have
\begin{eqnarray}
\nonumber
\langle q_{ijk} q_{rst}\rangle = \frac{1}{(4\pi)^2}\frac{1}{\sigma_i\sigma_j\sigma_k}\frac{1}{\sigma_r\sigma_s\sigma_t}\sum_{l_1l_2l_3}I_{l_1l_2l_3}^2 C_{l_1}C_{l_2}C_{l_3} \\
\nonumber
\times \{ \omega_{l_1}(R_i)\omega_{l_1}(R_r)\omega_{l_2}(R_j)\omega_{l_2}(R_s)\omega_{l_3}(R_k)\omega_{l_3}(R_t)\\
\nonumber
+ \omega_{l_1}(R_i)\omega_{l_1}(R_r)\omega_{l_2}(R_j)\omega_{l_2}(R_t)\omega_{l_3}(R_k)\omega_{l_3}(R_s)\\
\nonumber
+ \omega_{l_1}(R_i)\omega_{l_1}(R_s)\omega_{l_2}(R_j)\omega_{l_2}(R_r)\omega_{l_3}(R_k)\omega_{l_3}(R_t)\\
\nonumber
+ \omega_{l_1}(R_i)\omega_{l_1}(R_s)\omega_{l_2}(R_j)\omega_{l_2}(R_t)\omega_{l_3}(R_k)\omega_{l_3}(R_r)\\
\nonumber
+ \omega_{l_1}(R_i)\omega_{l_1}(R_t)\omega_{l_2}(R_j)\omega_{l_2}(R_r)\omega_{l_3}(R_k)\omega_{l_3}(R_s)\\
\nonumber
+ \omega_{l_1}(R_i)\omega_{l_1}(R_t)\omega_{l_2}(R_j)\omega_{l_2}(R_s)\omega_{l_3}(R_k)\omega_{l_3}(R_r)\} \\
\label{the_cov_qijk_stat}
\end{eqnarray}
The pixel properties are taken into account here by replacing
$C_{\ell}$ by $C_{\ell} \omega_{\ell}^{(pix)}
\omega_{\ell}^{(pix)}$. We have compared the analytical covariance
matrix obtained through Eq. \ref{the_cov_qijk_stat} with the
covariance matrix obtained with $10^4$ Gaussian simulations for the
same cosmological parameters defined in Subsect.
\ref{statistics_and_fnl}. The covariance matrix elements
$C_{ijk,rst}$, are compared by pairs in
Fig. \ref{the_analytical_sims_cov_matrix}, obtaining a very good
agreement.
\subsection{$f_{nl}$ Fisher matrix of the third order moments}
\label{fisher_wavelets}
We discuss the detectability of primary non-Gaussianity with the
third order moments. Assuming that the third order statistics are
Gaussian-like, we can use the Gaussian likelihood to constrain the
$f_{nl}$ parameter
\begin{equation}
L(f_{nl}) = C_0 e^{-\chi^2(f_{nl})/2},
\label{likelihood}
\end{equation}
where $C_0$ is a constant and $\chi^2(f_{nl})$ is given by
\begin{equation}
\chi^2(f_{nl}) = \sum_{ijk,rst}(q_{ijk}^{obs}-\langle q_{ijk}
\rangle_{f_{nl}})C^{-1}_{ijk,rst}(q_{rst}^{obs}-\langle q_{rst}
\rangle_{f_{nl}}).
\label{chi_analytic}
\end{equation}
$C^{-1}_{ijk,rst}$ is the inverse of the covariance matrix of the
third order statistics that we have computed analytically in the
previous section, $q_{ijk}^{obs}$ are the third order statistics
obtained from the data and $\langle q_{ijk} \rangle_{f_{nl}}$ are the
expected values of the third order statistics for a given model with
$f_{nl}$. As we have seen in Subsect. \ref{statistics_and_fnl},
$\langle q_{ijk} \rangle_{f_{nl}} = f_{nl} \alpha_{ijk}$, with
$\alpha_{ijk}$ a constant independent of $f_{nl}$. Using this on
Eq. \ref{chi_analytic}, we have
\begin{equation}
\chi^2(f_{nl}) =
\sum_{ijk,rst}(q_{ijk}^{obs}-\alpha_{ijk}f_{nl})C^{-1}_{ijk,rst}(q_{rst}^{obs}-\alpha_{rst}f_{nl}).
\label{chi_analytic_fnl}
\end{equation}
The variance of the $f_{nl}$ parameter can be computed using the
Fisher matrix, which leads to: 
\begin{eqnarray}
\nonumber
\sigma^2(f_{nl}) = \frac{-1}{\frac{\partial^2 log L(f_{nl})}{\partial
    f_{nl}^2}} = \frac{1}{\frac{1}{2}\frac{\partial^2
    \chi^2(f_{nl})}{\partial f_{nl}^2}} = \\
= \frac{1}{\sum_{ijk,rst}\alpha_{ijk}C^{-1}_{ijk,rst}\alpha_{rst}}.
\label{sigma_fnl_likelihood}
\end{eqnarray}
\subsection{Principal Component Analysis of the third order moments}
\label{pca_subsection}
One of the most significant advantages of the wavelet-based analysis
for this type of non-Gaussianity is that we are able to reduce the
non-Gaussian information present in the bispectrum
$b_{\ell_1\ell_2\ell_3}$ (about $10^7$ elements) to a set of several
hundreds of $q_{ijk}$ statistics. However, these quantities are
correlated and in certain conditions their covariance matrix may be
ill-conditioned. There are different approaches to deal with these
matrices. We will apply a Principal Component Analysis (PCA) in all
the tests where the wavelet coefficient covariance matrix is involved
in order to keep the errors related to the covariance matrix under a
certain threshold.

The $\chi^2$ statistic of a random variable ${\bf x}$ of dimension $n$
can be written
\begin{equation}
\chi^2 = \sum_{i,j=1}^n x_i C_{ij}^{-1} x_j,
\end{equation}
where $C$ is the covariance matrix of this variable. For any positive
definite covariance matrix, it is possible to find a linear
transformation of the ${\bf x}$ vector where the corresponding
covariance matrix is the identity
\begin{equation}
\chi^2 = \sum_{i=1}^n y_i^2
\end{equation}
where $y_i = \sum_{j=1}^n(D^{1/2}R^t)_{ij}x_j$ and $D$ and $R$ are the
eigenvalue and eigenvector matrices of the covariance matrix, $C =
RDR^t$. In particular we sort the eigenvalues in descending order $D_i
> D_{i+1}$. We can define a partial $\chi^2_m$ statistic
\begin{equation}
\chi_m^2 = \sum_{i=1}^m y_i^2
\end{equation}
such that only the $y_i$ quantities related to the largest $m$
eigenvalues are considered. The Fisher matrix for the $f_{nl}$
parameter using $\chi_m^2$ is defined as 
\begin{equation}
\sigma_m^2(f_{nl}) \equiv \frac{1}{\frac{1}{2}\frac{\partial \chi^2_m(f_{nl})}{\partial f_{nl}^2}}.
\label{sigma_fnl_likelihood_pca1}
\end{equation}
As the third order moments are linearly proportional to $f_{nl}$, $x_j
= q_j - f_{nl} \alpha_j$ and defining $\beta_i = \sum_{j=1}^n
(D^{-1/2}R^t)_{ij} \alpha_j$, the Fisher matrix for the $f_{nl}$
parameter is
\begin{equation}
\sigma_m^2(f_{nl}) = \frac{1}{\sum_{i=1}^m\beta_i^2}.
\label{sigma_fnl_likelihood_pca2}
\end{equation}
\subsection{Bispectrum estimator and error bars}
The bispectrum-based estimators are the most widely applied tools for
detecting primordial non-Gaussianity. Considering the angle averaged
bispectrum $B_{\ell_1\ell_2\ell_3}$ as defined in \citet{komatsu2001},
it can be shown that the unbiased bispectrum-based minimum variance
estimator for the full-sky limit and homogeneous noise is
\citep{creminelli2006}
\begin{eqnarray}
\nonumber
\hat{f}_{nl} = \frac{1}{N}\sum_{\ell_i,m_i} \left ( \begin{array}{ccc}
  \ell_1 & \ell_2 & \ell_3 \\ m_1 & m_2 & m_3 \end{array}\right )
\frac{B^{model}_{\ell_1 \ell_2 \ell_3}}{C_{\ell_1}C_{\ell_2}C_{\ell_3}} \\
\times a_{\ell_1m_1}a_{\ell_2m_2}a_{\ell_3m_3}.
\end{eqnarray}
The unbiased bispectrum-based minimum variance estimator for the
incomplete sky limit and inhomogeneous noise is \citep{creminelli2006} 
\begin{eqnarray}
\nonumber 
\hat{f}_{nl} = \frac{1}{N}\sum_{\ell_i,m_i} \Big ( \big \langle
a_{\ell_1m_1}a_{\ell_2m_2}a_{\ell_3m_3}
\big \rangle_{f_{nl}=1} \\
\nonumber
\times C^{-1}_{\ell_1m_1,\ell_4m_4}C^{-1}_{\ell_2m_2,\ell_5m_5}C^{-1}_{\ell_3m_3,\ell_6m_6} \\
\nonumber
\times a_{\ell_4m_4}a_{\ell_5m_5}a_{\ell_6m_6} \\
\nonumber 
-3 \big \langle a_{\ell_1m_1}a_{\ell_2m_2}a_{\ell_3m_3}\big \rangle_{f_{nl}=1} \\
C^{-1}_{\ell_1m_1,\ell_2m_2}C^{-1}_{\ell_3m_3,\ell_4m_4}a_{\ell_4m_4}
\Big ).
\end{eqnarray}
In particular, the signal-to-noise ratio for the $f_{nl}$ parameter,
using a Fisher analysis for a full-sky and homogeneous noise
experiment is \citep{komatsu2001}
\begin{equation}
\frac{f_{nl}}{\sigma^2(f_{nl})} = \sum_{2 \le \ell_1 \le \ell_2 \le
  \ell_3}\frac{B^{model}_{\ell_1 \ell_2 \ell_3}B^{model}_{\ell_1
    \ell_2 \ell_3}}{\sigma^2_{\ell_1 \ell_2 \ell_3}},
\label{sigma_fnl_bispectrum_intro}
\end{equation}
where $\sigma^2_{\ell_1 \ell_2 \ell_3}$ is the variance of the
bispectrum. 

Given a map with $N_{pix}$ pixels, the number of operations to obtain
the local $f_{nl}$ through the KSW algorithm based on the bispectrum
is $\sim 100 N_{pix}^{3/2}$ \citep{komatsu2005}. Our algorithm based
on the SMHW requires $\sim n_{scal} N_{pix}^{3/2}$ operations, where
$n_{scal}$ is the considered number of scales ($n_{scal} \sim 15$ for
a WMAP-like experiment). Therefore, the wavelet algorithm is about an
order of magnitude faster than the bispectrum algorithm for this kind
of non-Gaussianity.
\section{Comparison with the optimal bispectrum estimator}
\label{comparison_with_opt_bisp_est}
In this Section we study the values of $\sigma(f_{nl})$ obtained
with the wavelet and bispectrum estimators for a full-sky and
homogeneous noise experiment and for WMAP-like 5-year and 7-year V+W
combined maps.
\subsection{Estimated error bars for $f_{nl}$ with an ideal experiment}
\begin{figure}
  \center
  \includegraphics[height=5.0cm,width=8.0cm]{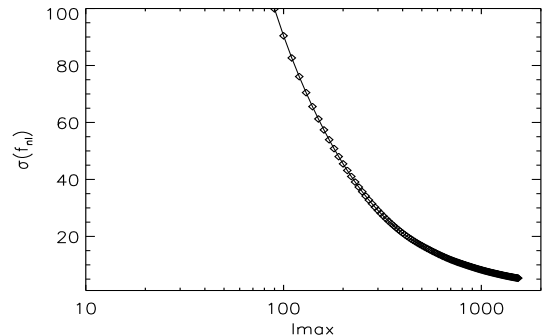}
  \caption{$\sigma(f_{nl})$ for different $\ell_{max}$ using the
    Fisher matrix of the bispectrum.\label{sigma_fnl_bispectrum}}
\end{figure}
We consider the local primordial bispectrum of a noiseless experiment
with an angular resolution of 6.9 arc minutes ($\ell_{max} = 1535$),
and a cosmological model characterised by $\Omega_{cdm}=0.25$,
$\Omega_{b}=0.05$, $\Omega_{\Lambda}=0.70$, $\tau = 0.09$, $h=0.73$
and $n=1$. The variance of $f_{nl}$ is computed through the Fisher
matrix of the bispectrum (see
Eq. \ref{sigma_fnl_bispectrum_intro}). We have used the gTfast code to
estimate the radiation transfer function and evaluate the expected
values of the local bispectrum for this experiment following Eq. (19)
of \citet{komatsu2001}. We have computed $\sigma(f_{nl})$ for
different $\ell_3 \le \ell_{max}$ (see Fig.
\ref{sigma_fnl_bispectrum}).
\begin{figure*}
\begin{center}
\includegraphics[width=3.2cm] {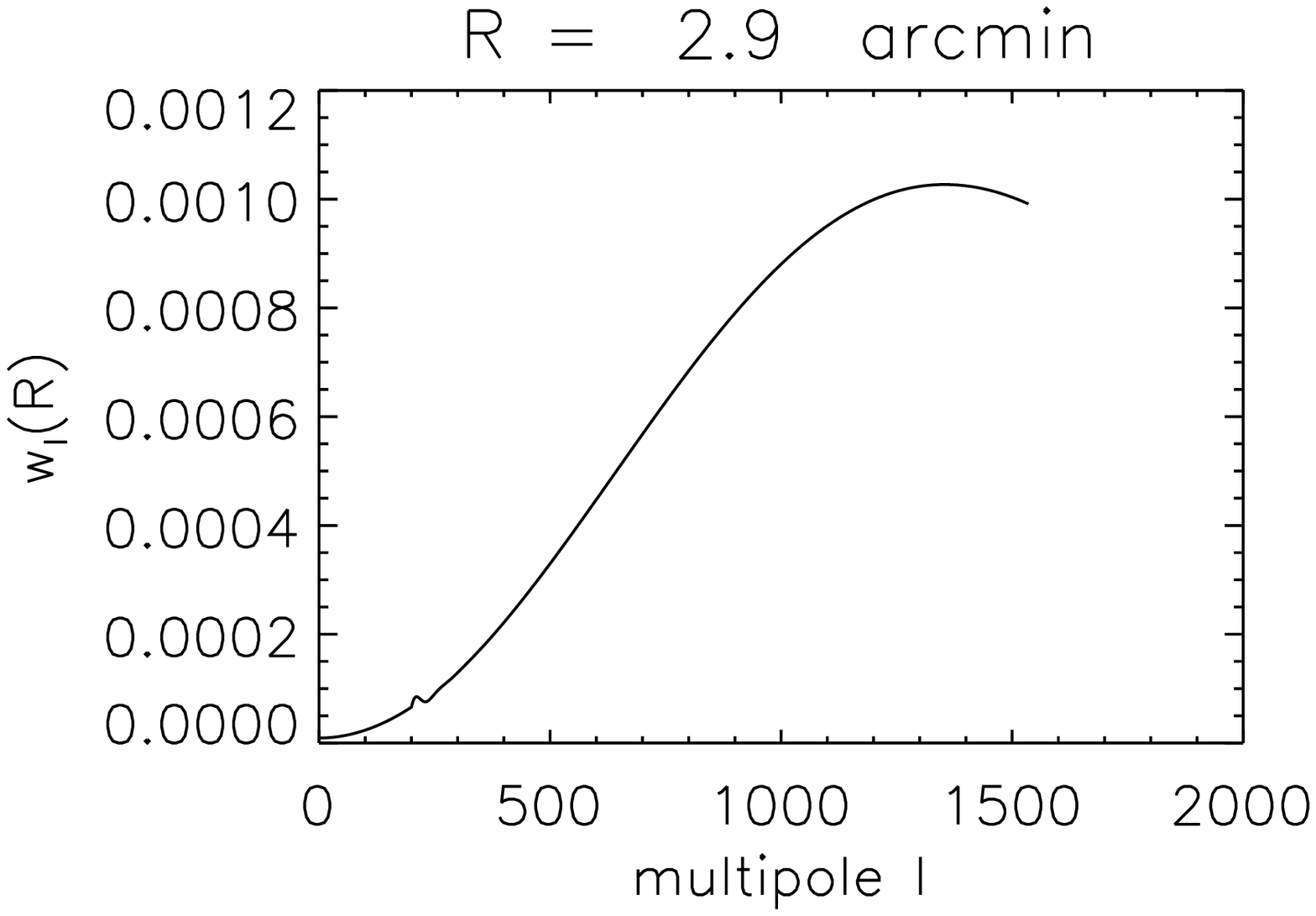}
\includegraphics[width=3.2cm] {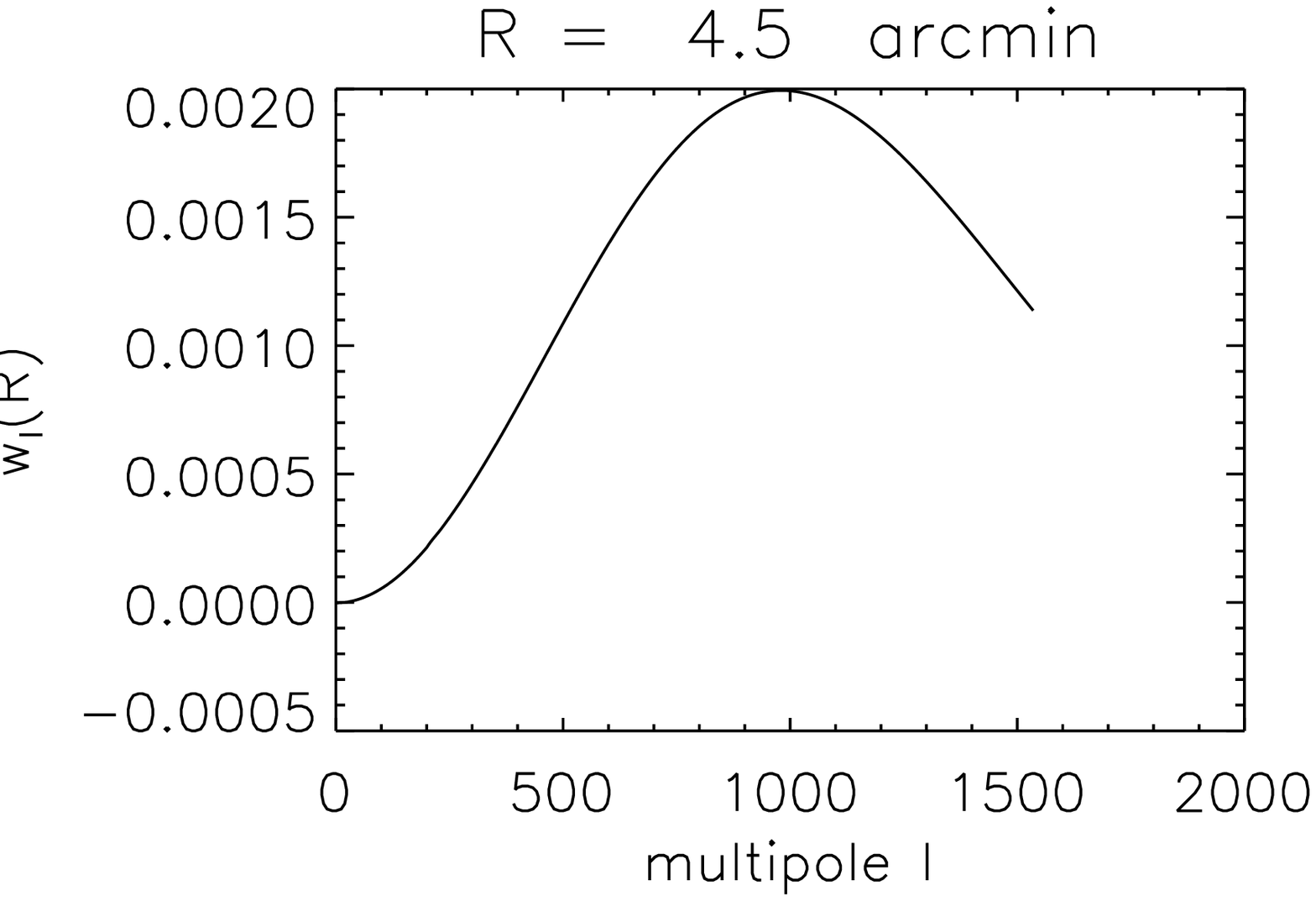}
\includegraphics[width=3.2cm] {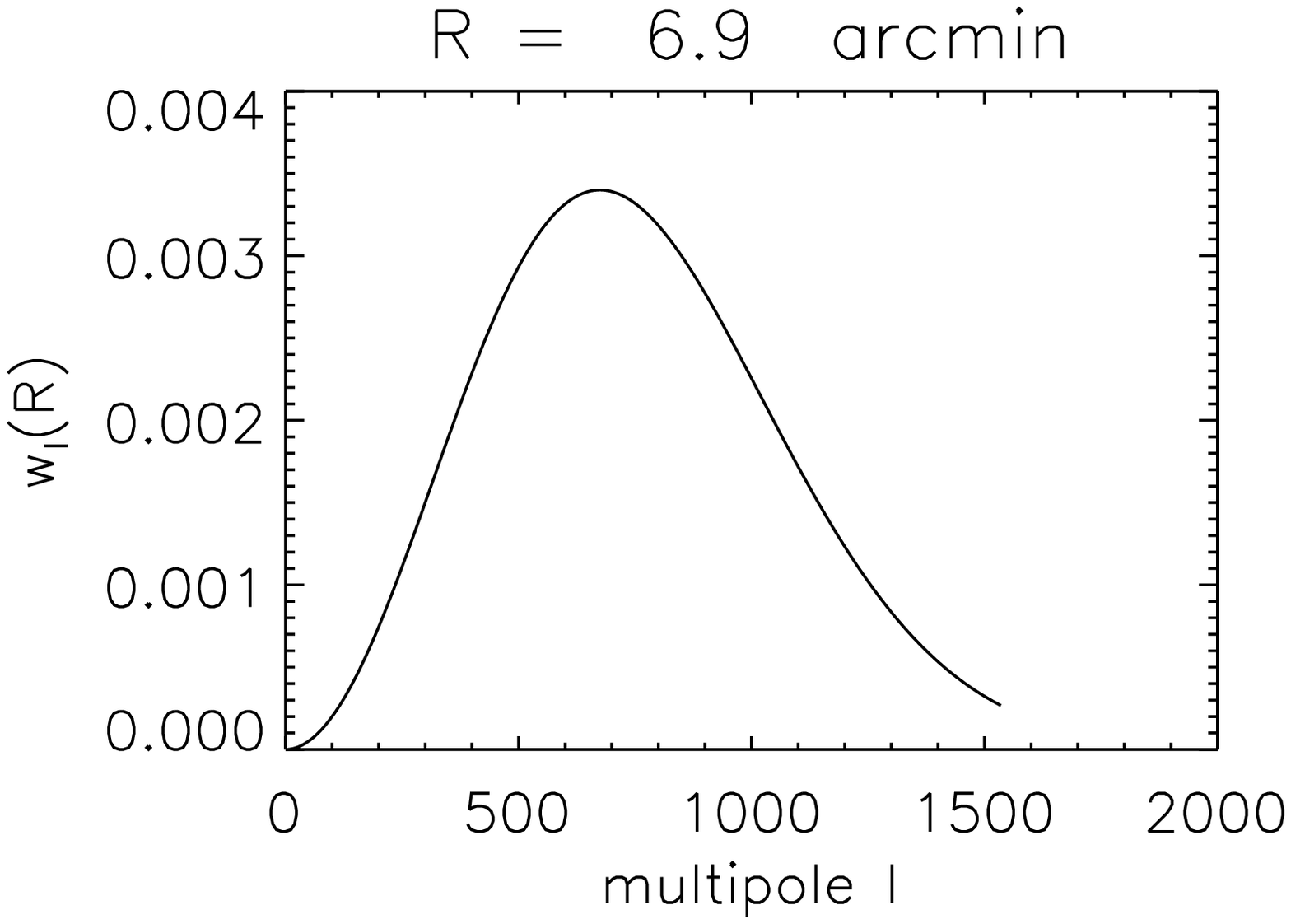}
\includegraphics[width=3.2cm] {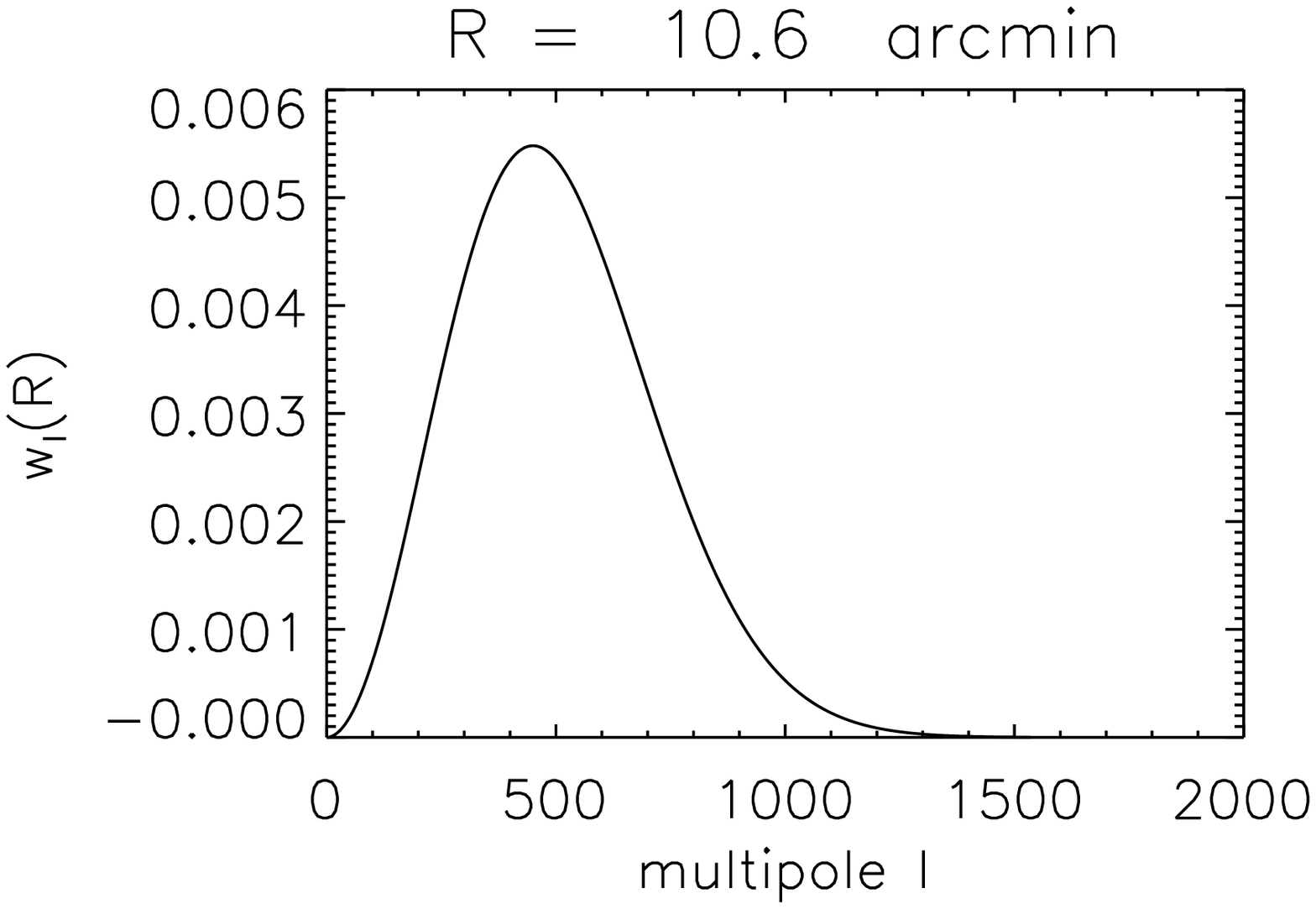}
\includegraphics[width=3.2cm] {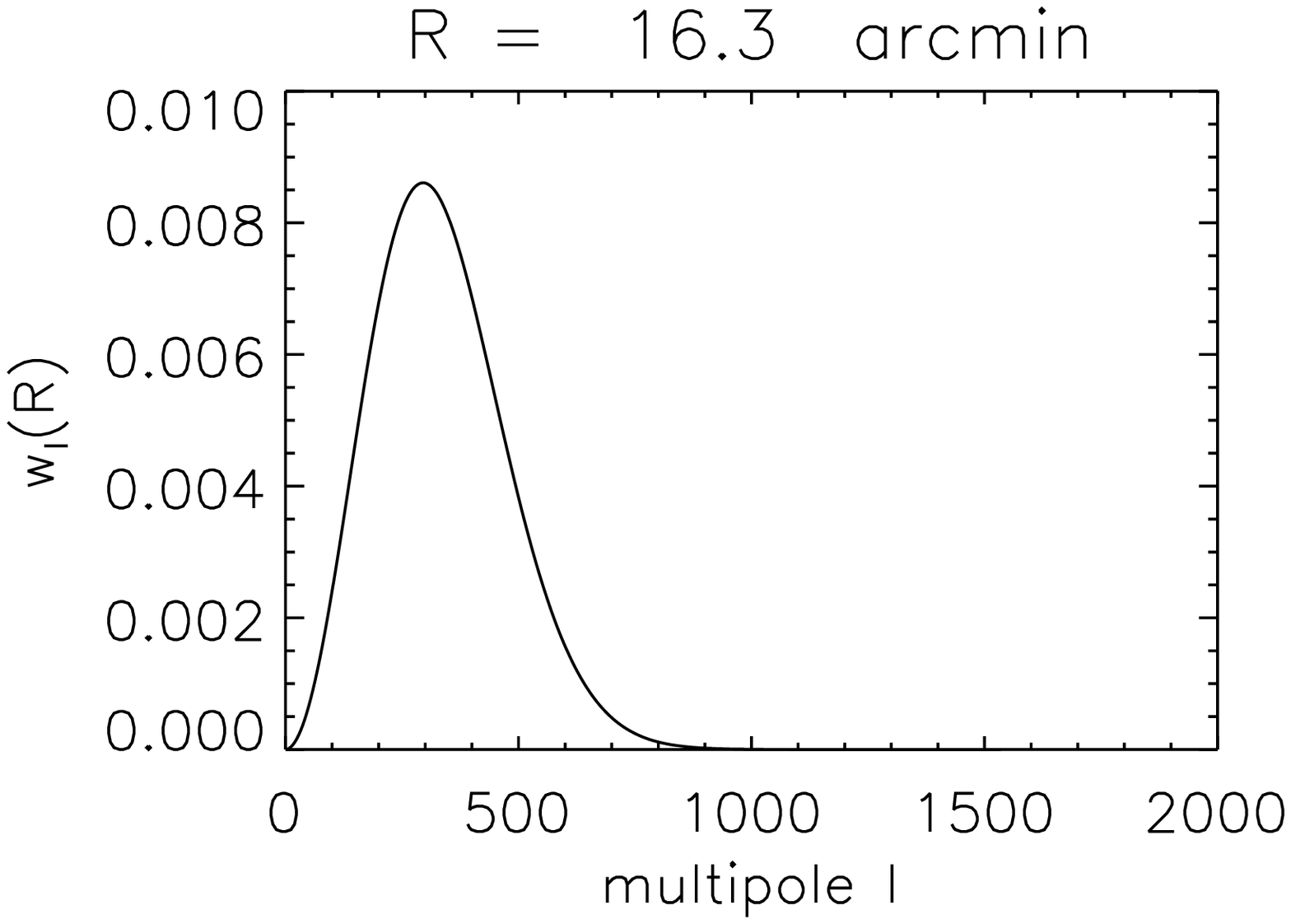}
\includegraphics[width=3.2cm] {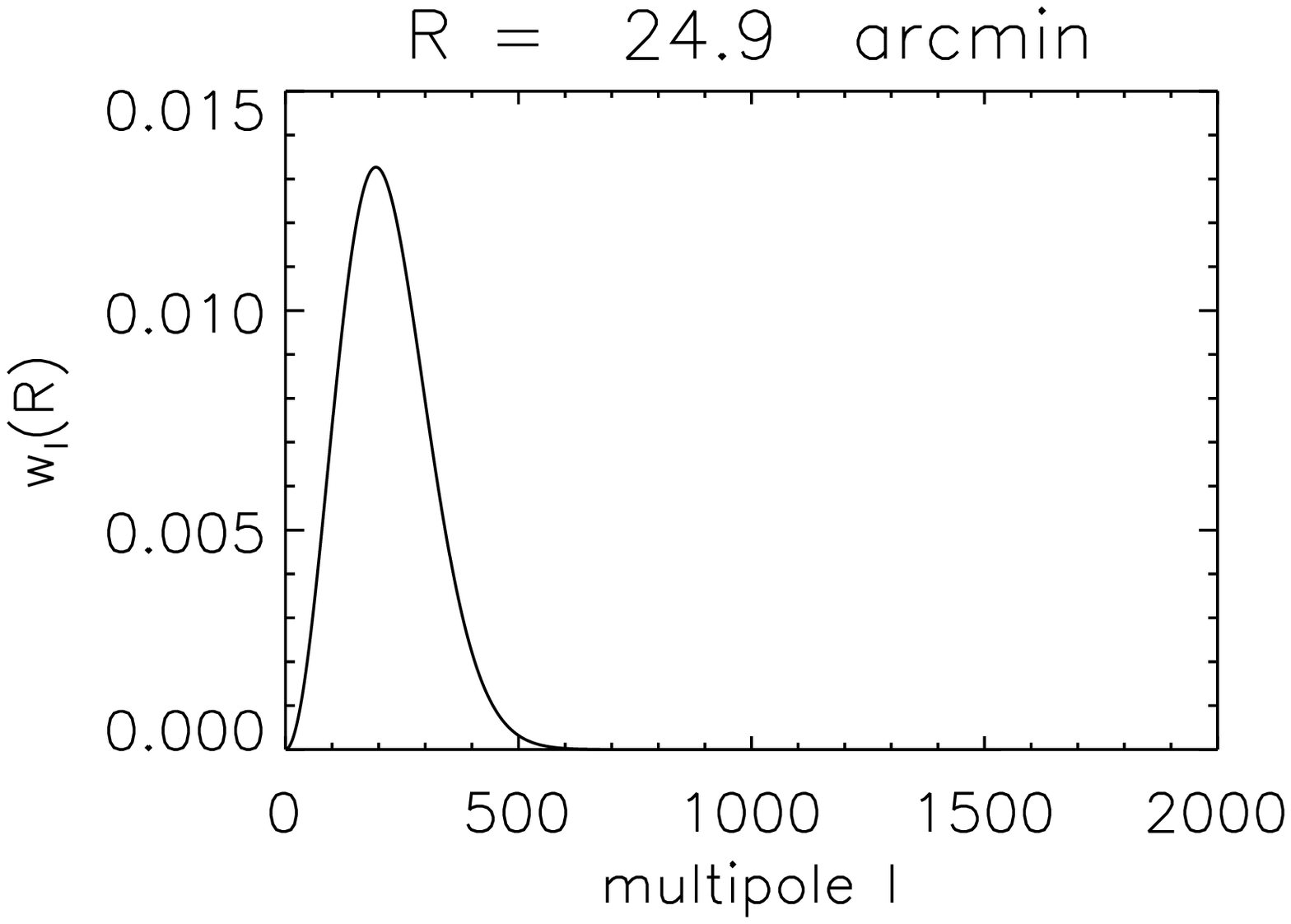}
\includegraphics[width=3.2cm] {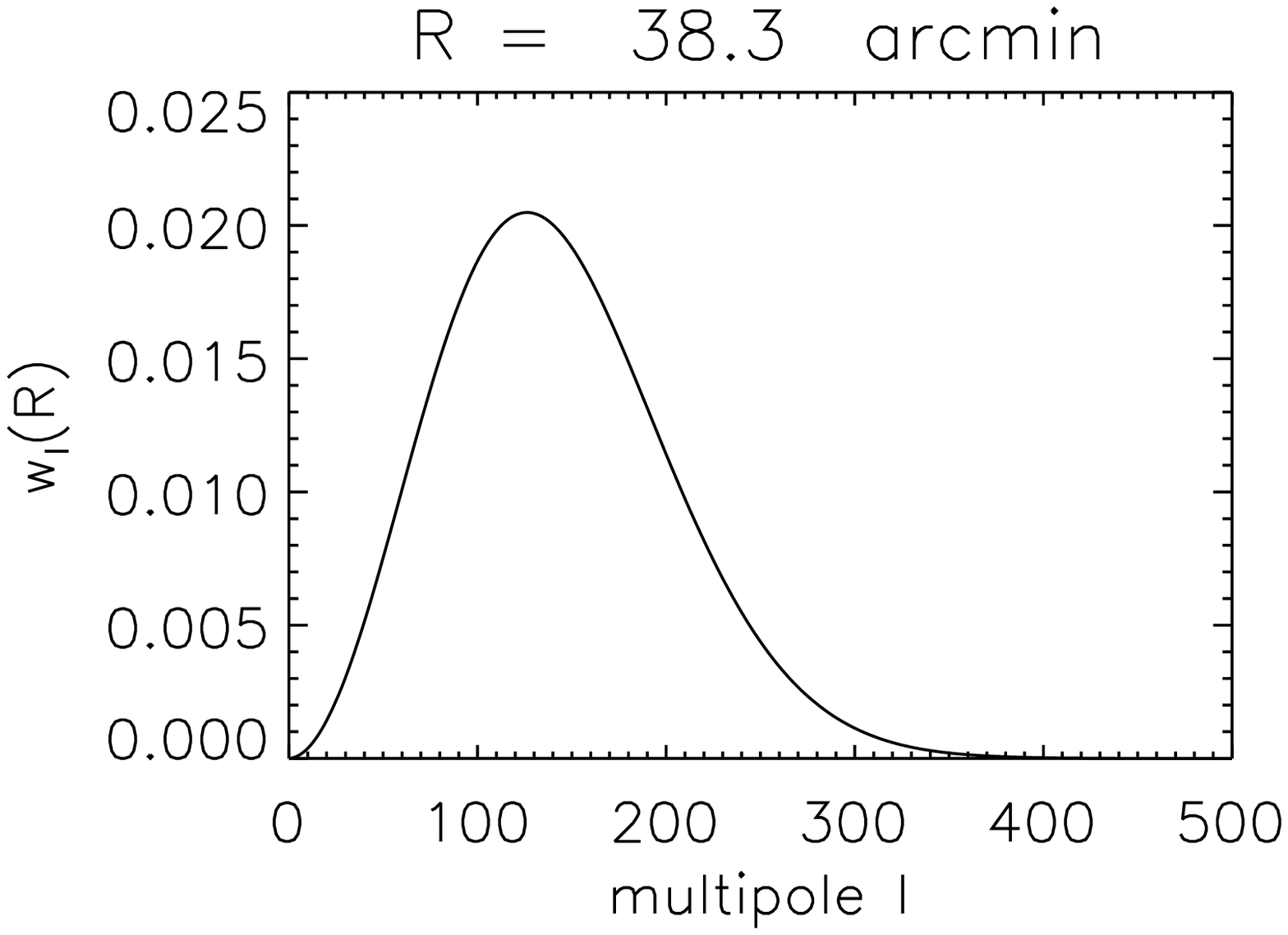}
\includegraphics[width=3.2cm] {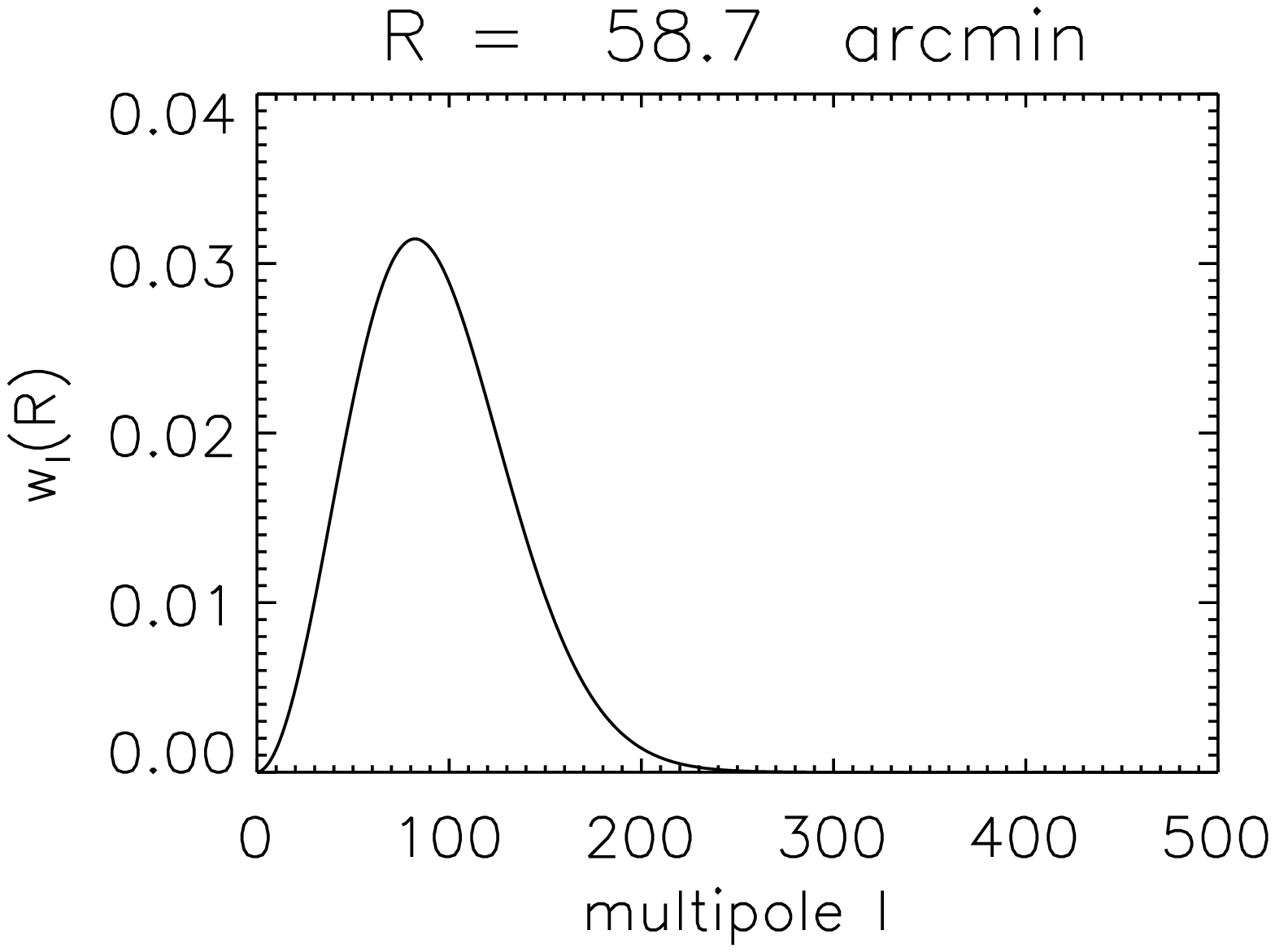}
\includegraphics[width=3.2cm] {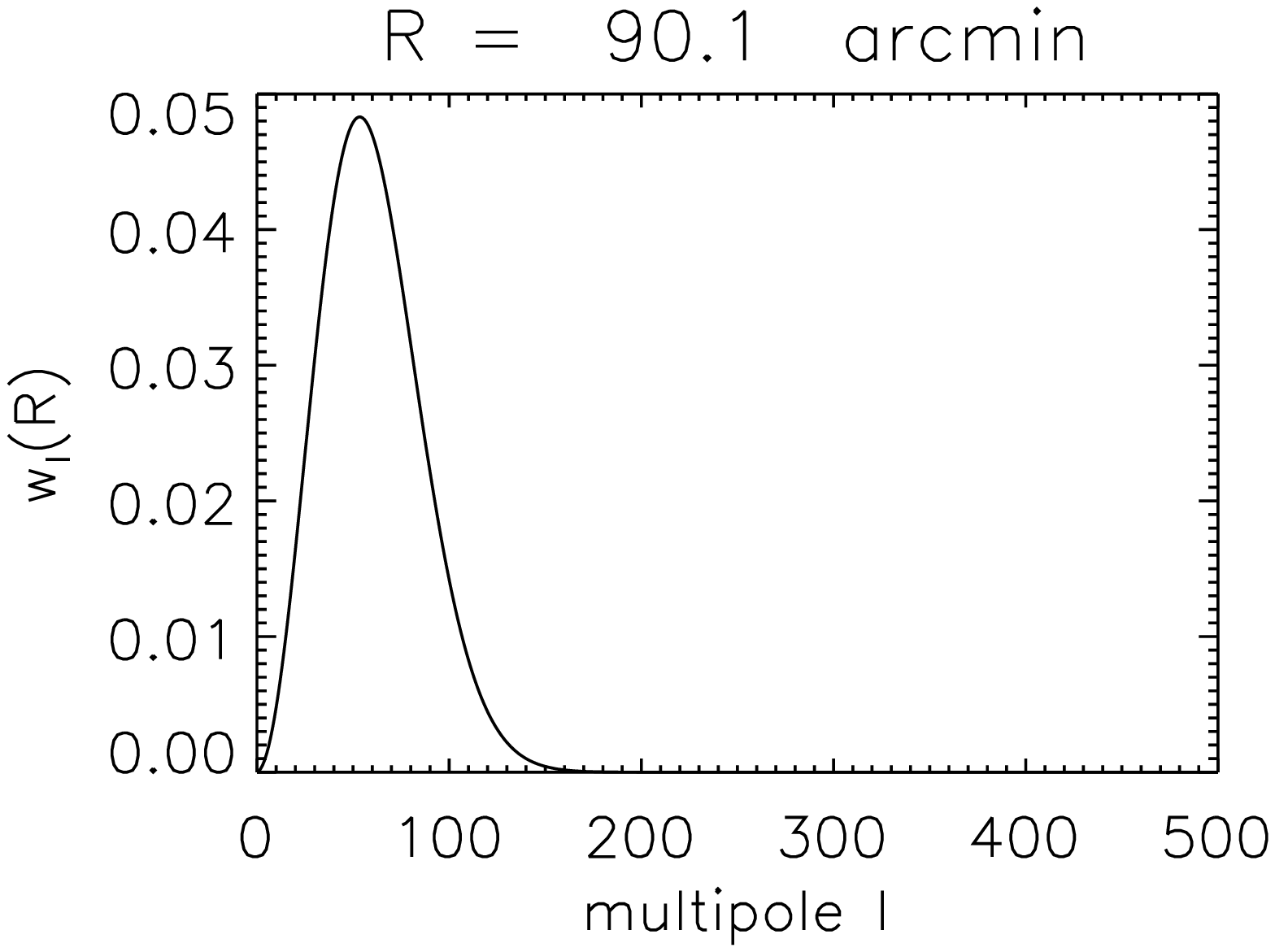}
\includegraphics[width=3.2cm] {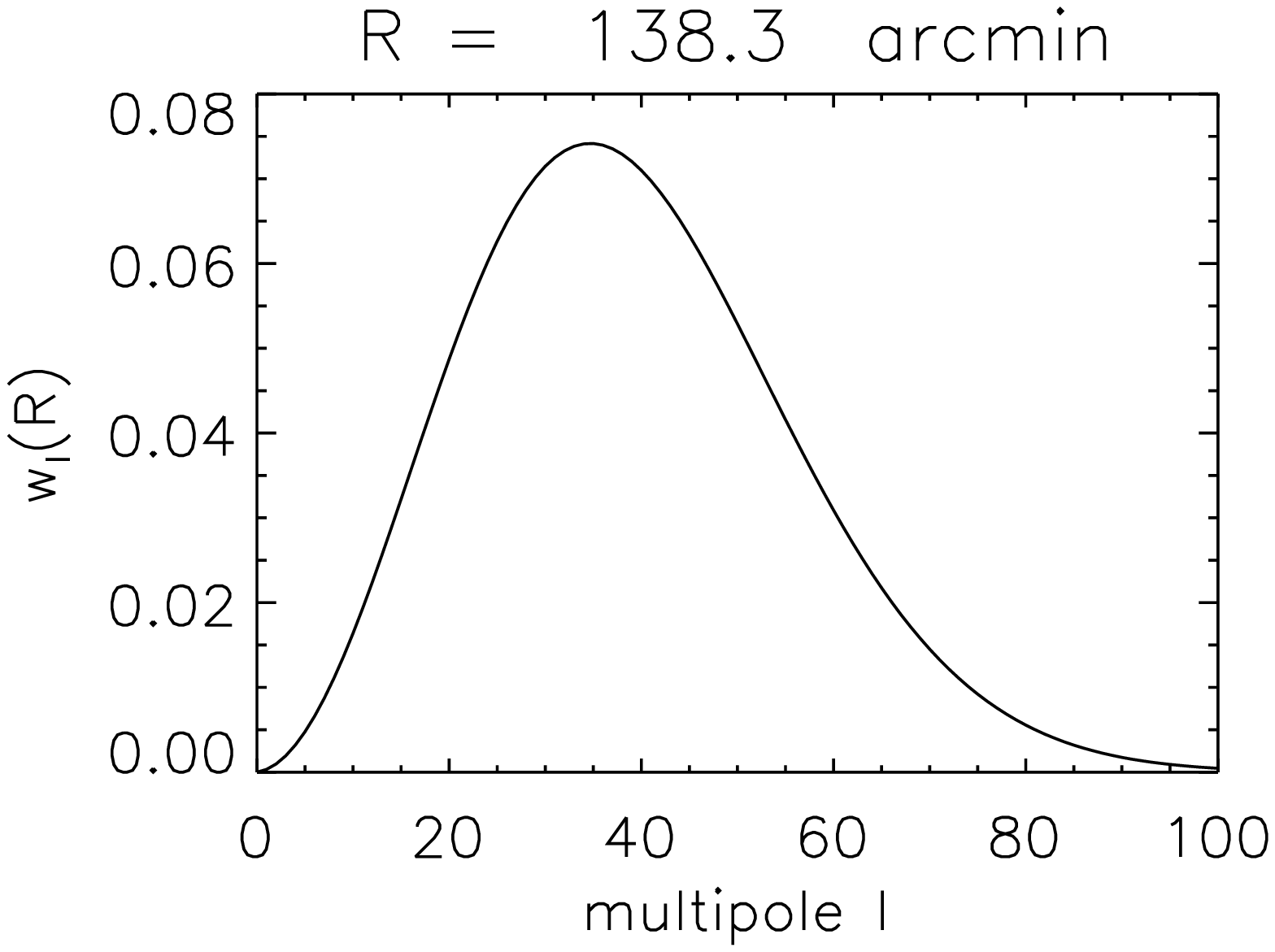}
\includegraphics[width=3.2cm] {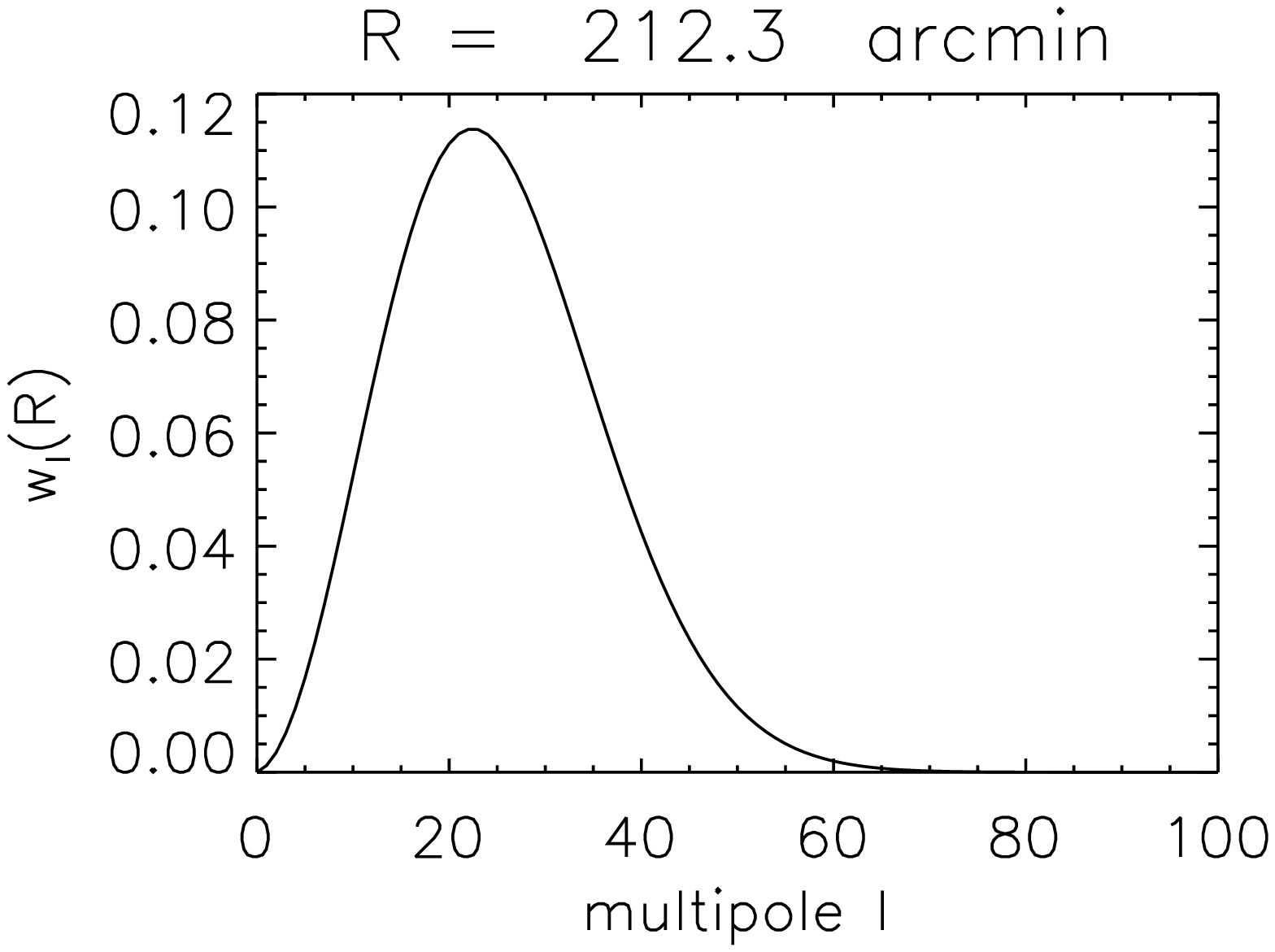}
\includegraphics[width=3.2cm] {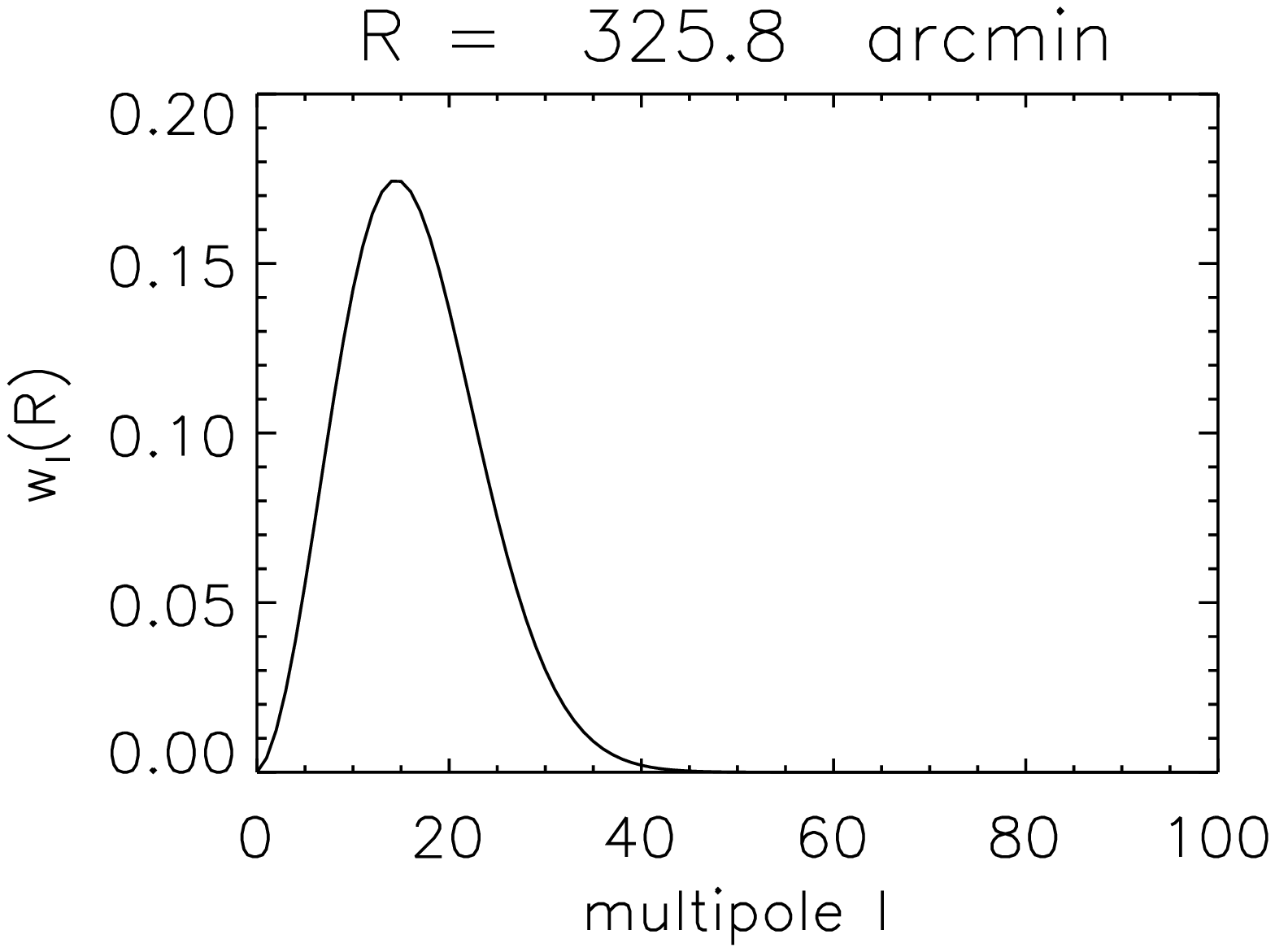}
\includegraphics[width=3.2cm] {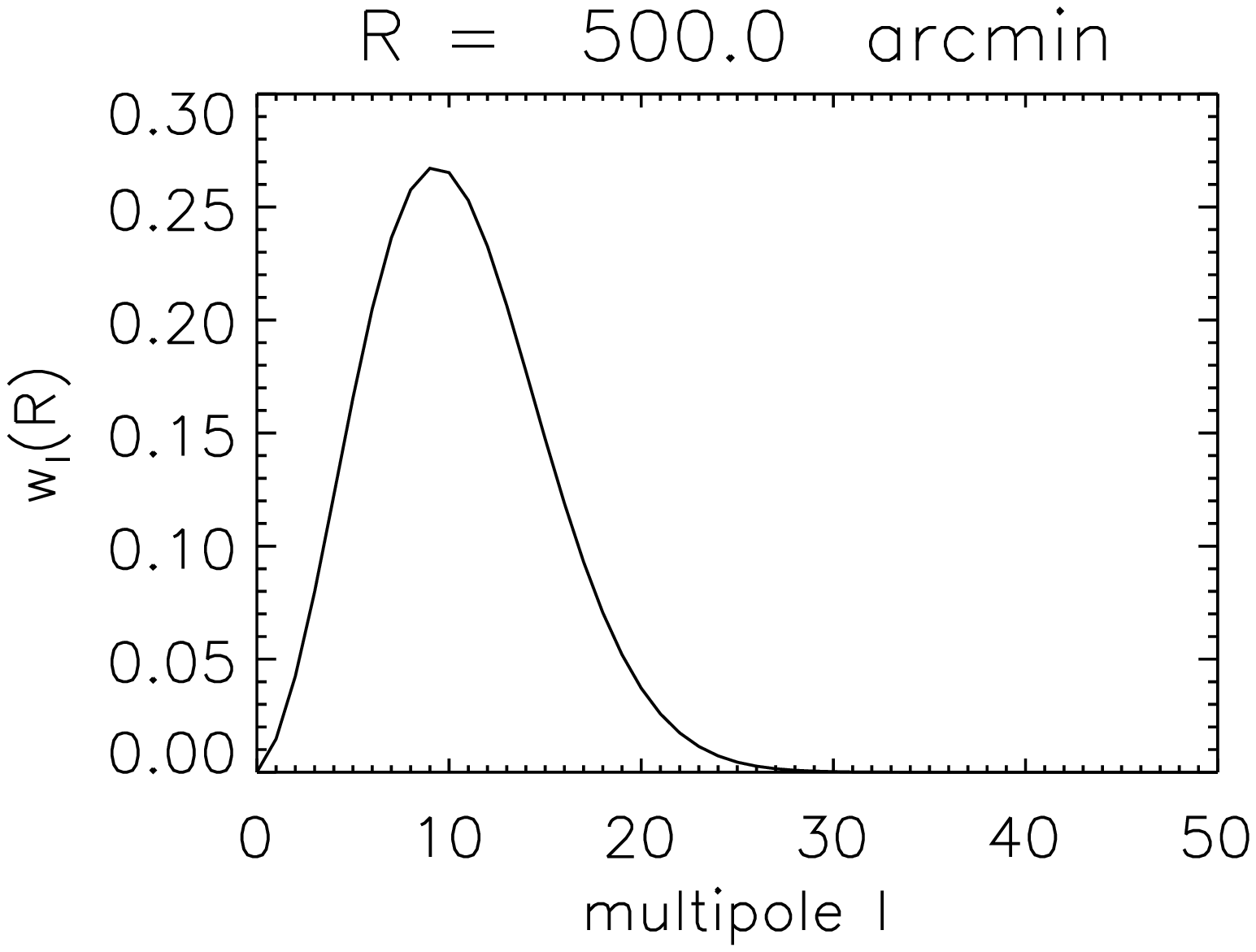}
\includegraphics[width=3.2cm] {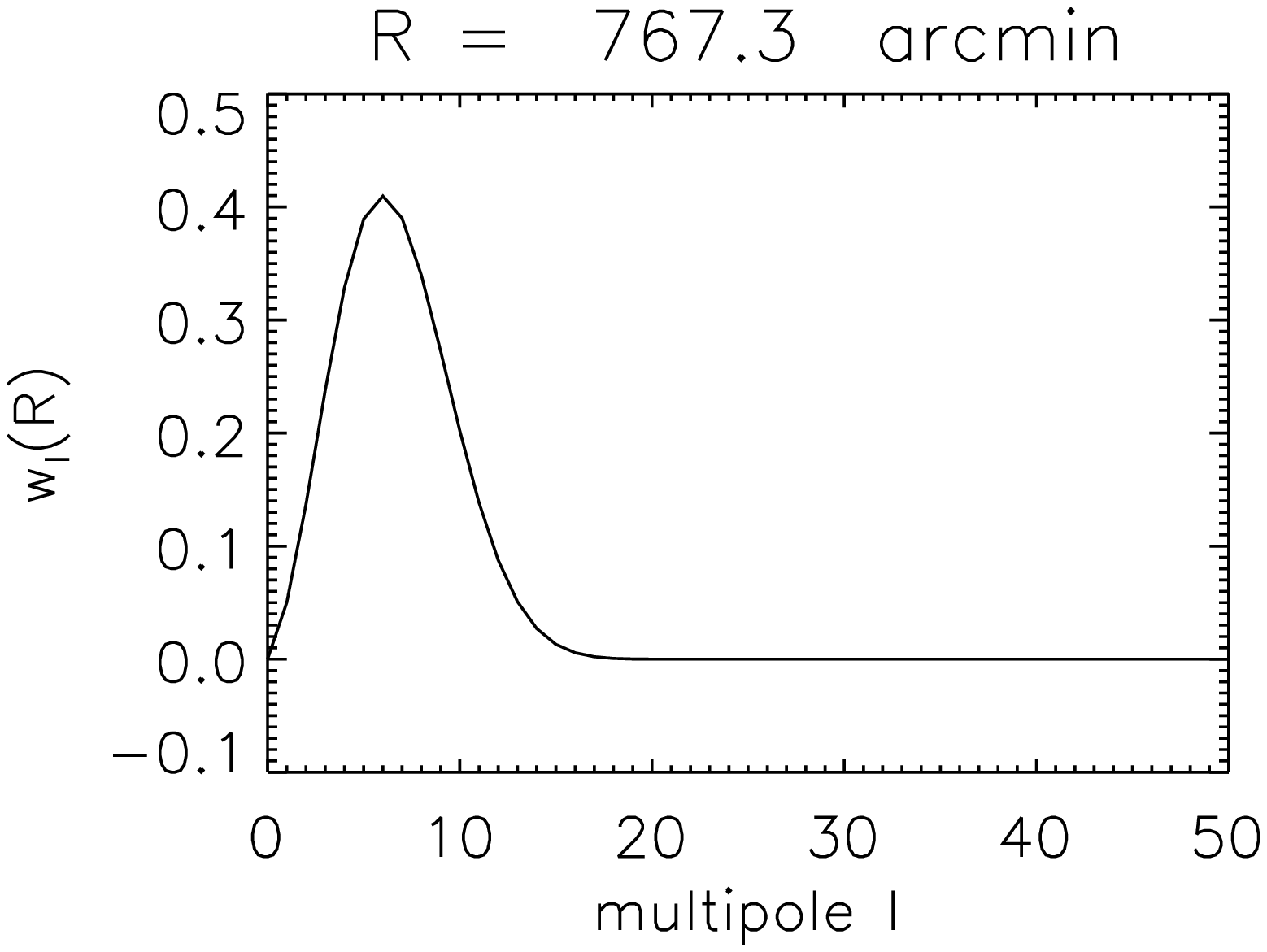}
\includegraphics[width=3.2cm] {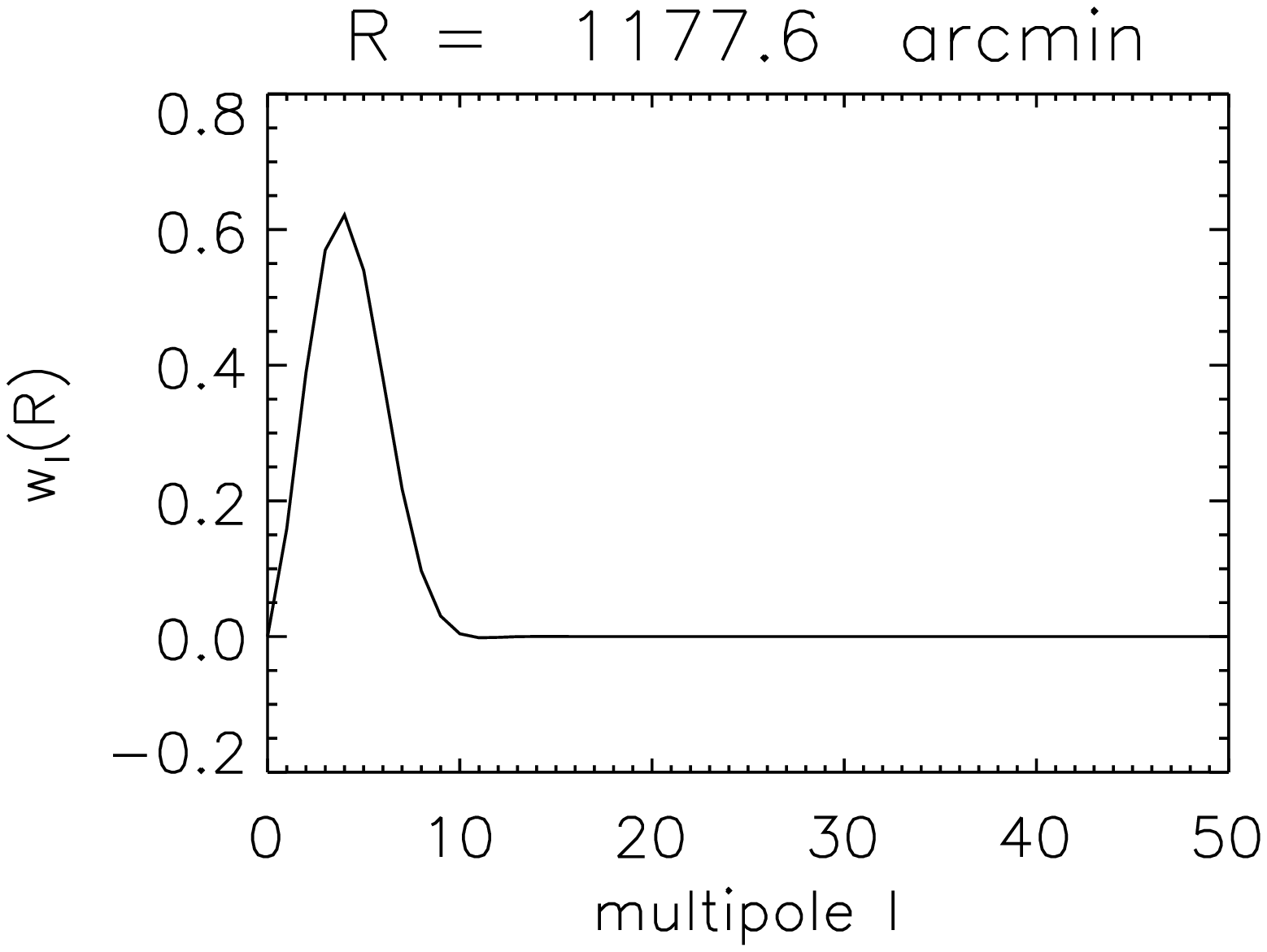}
\includegraphics[width=3.2cm] {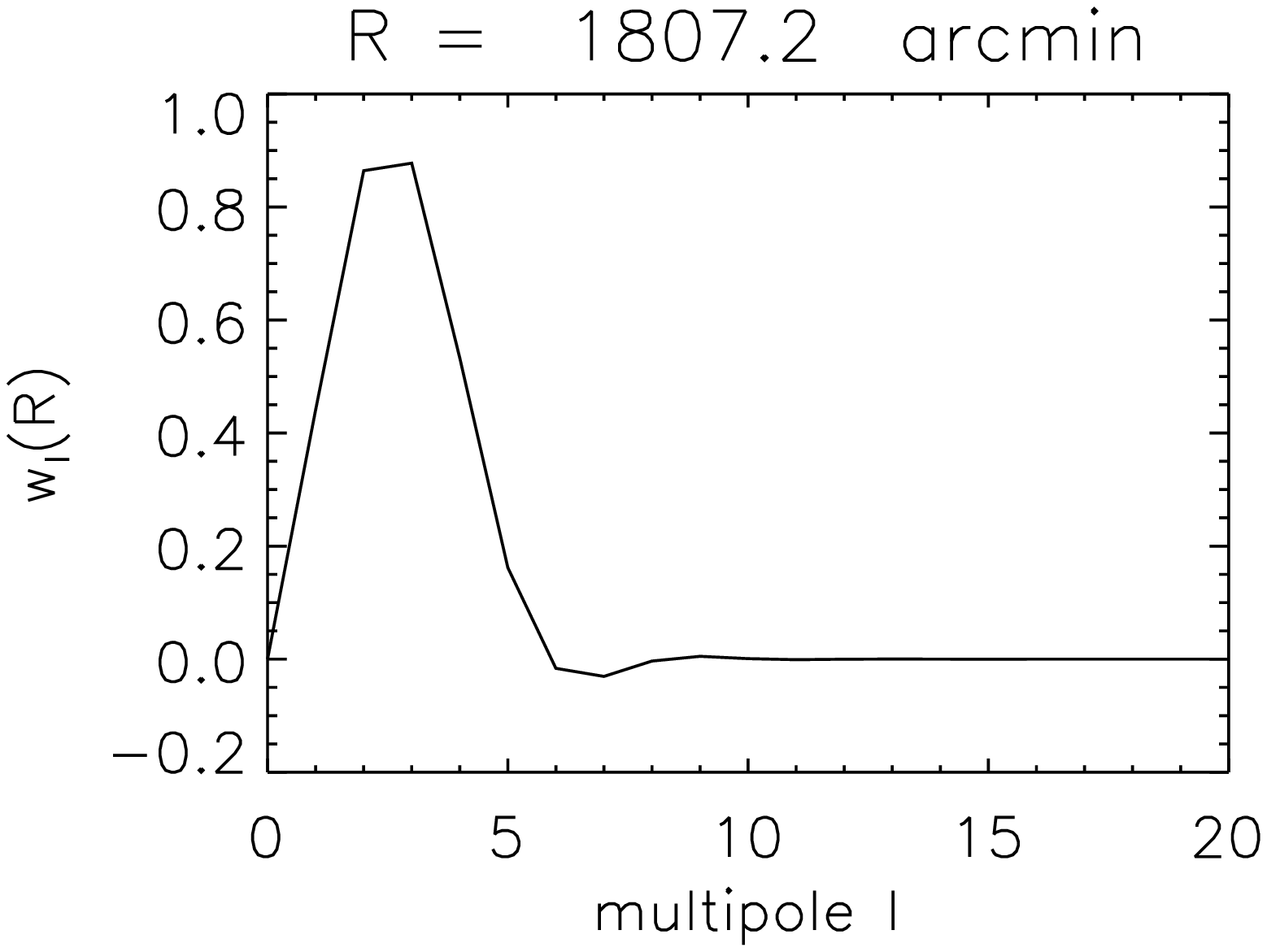}
\includegraphics[width=3.2cm] {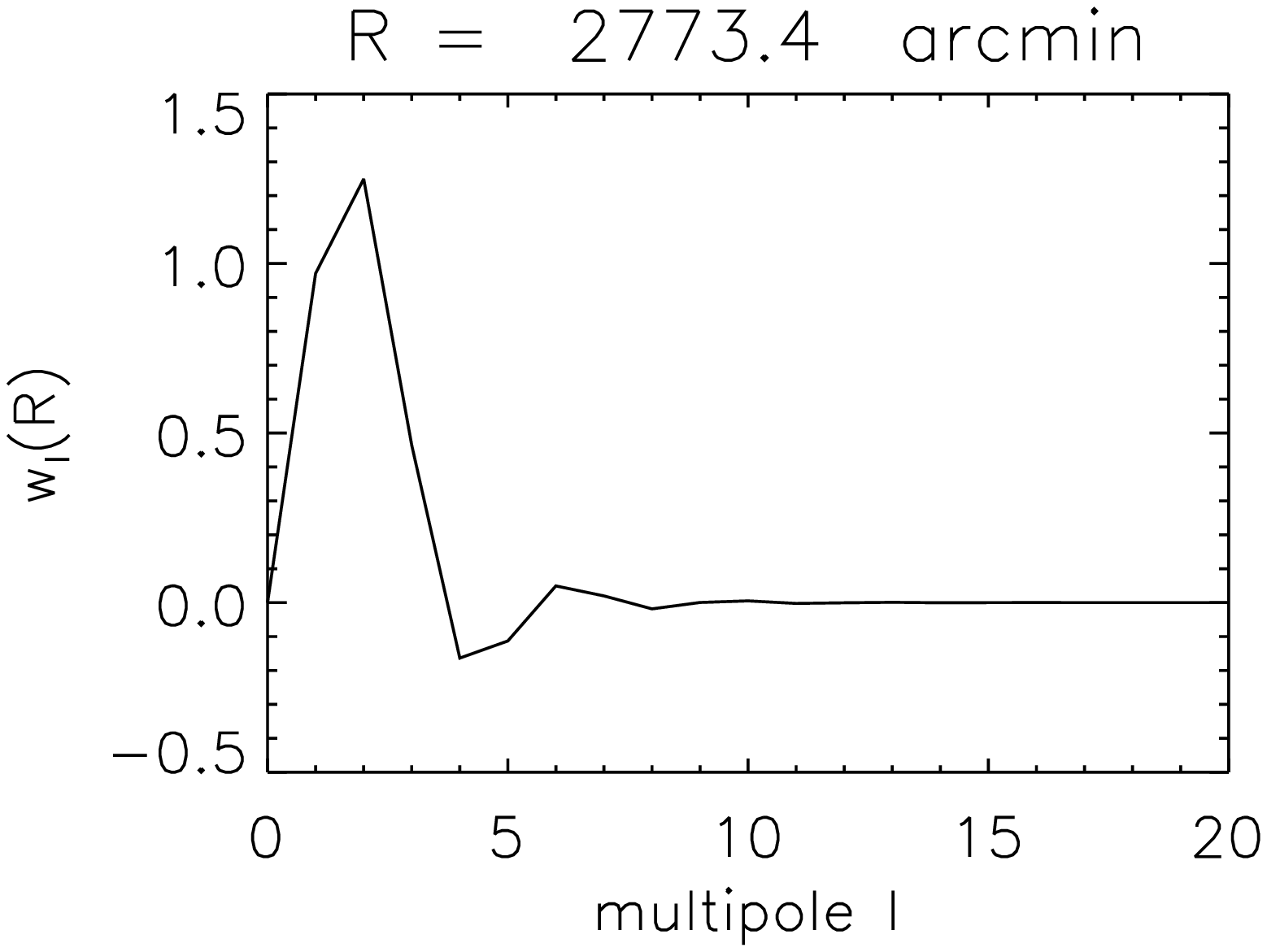}
\includegraphics[width=3.2cm] {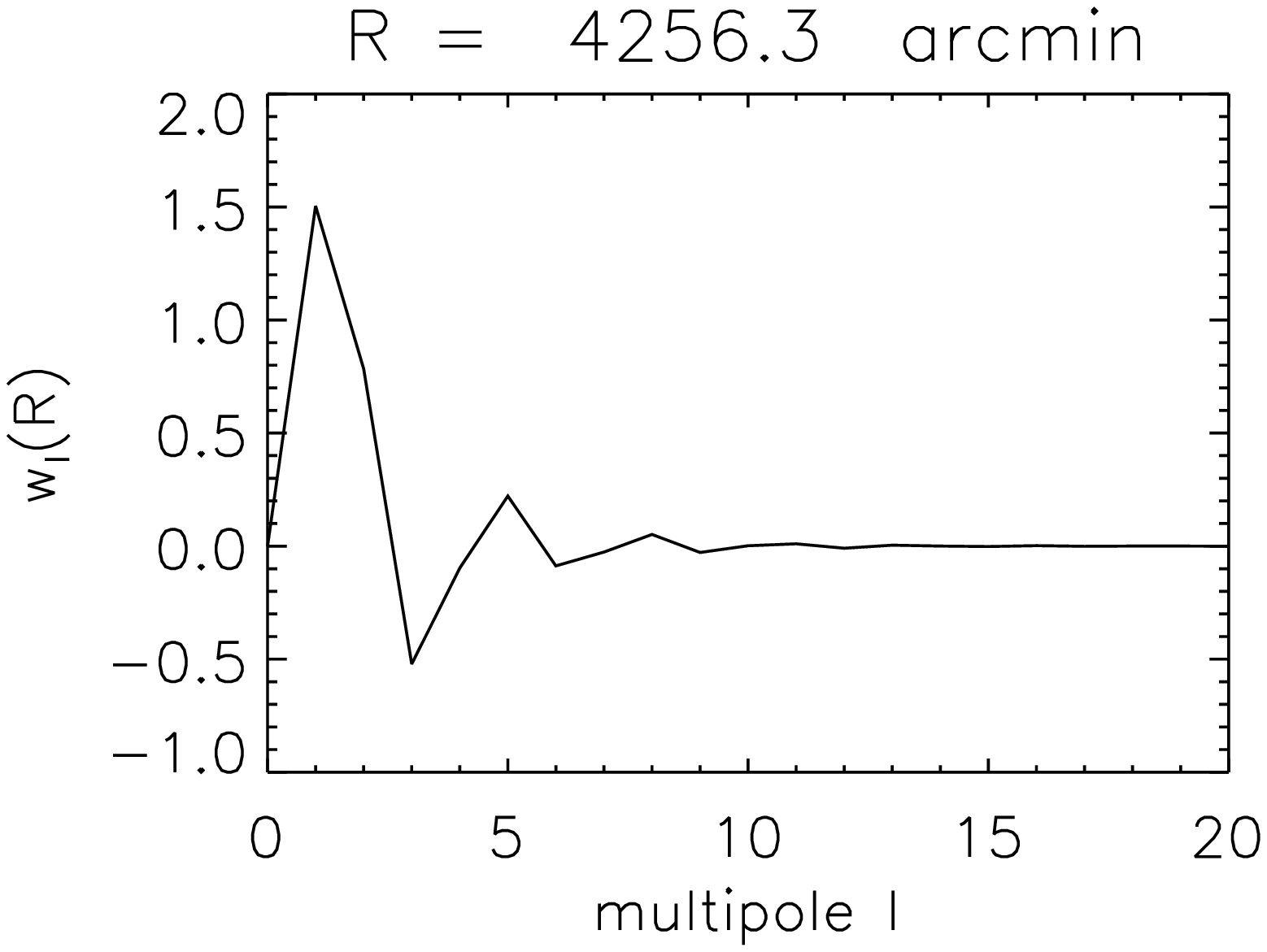}
\includegraphics[width=3.2cm] {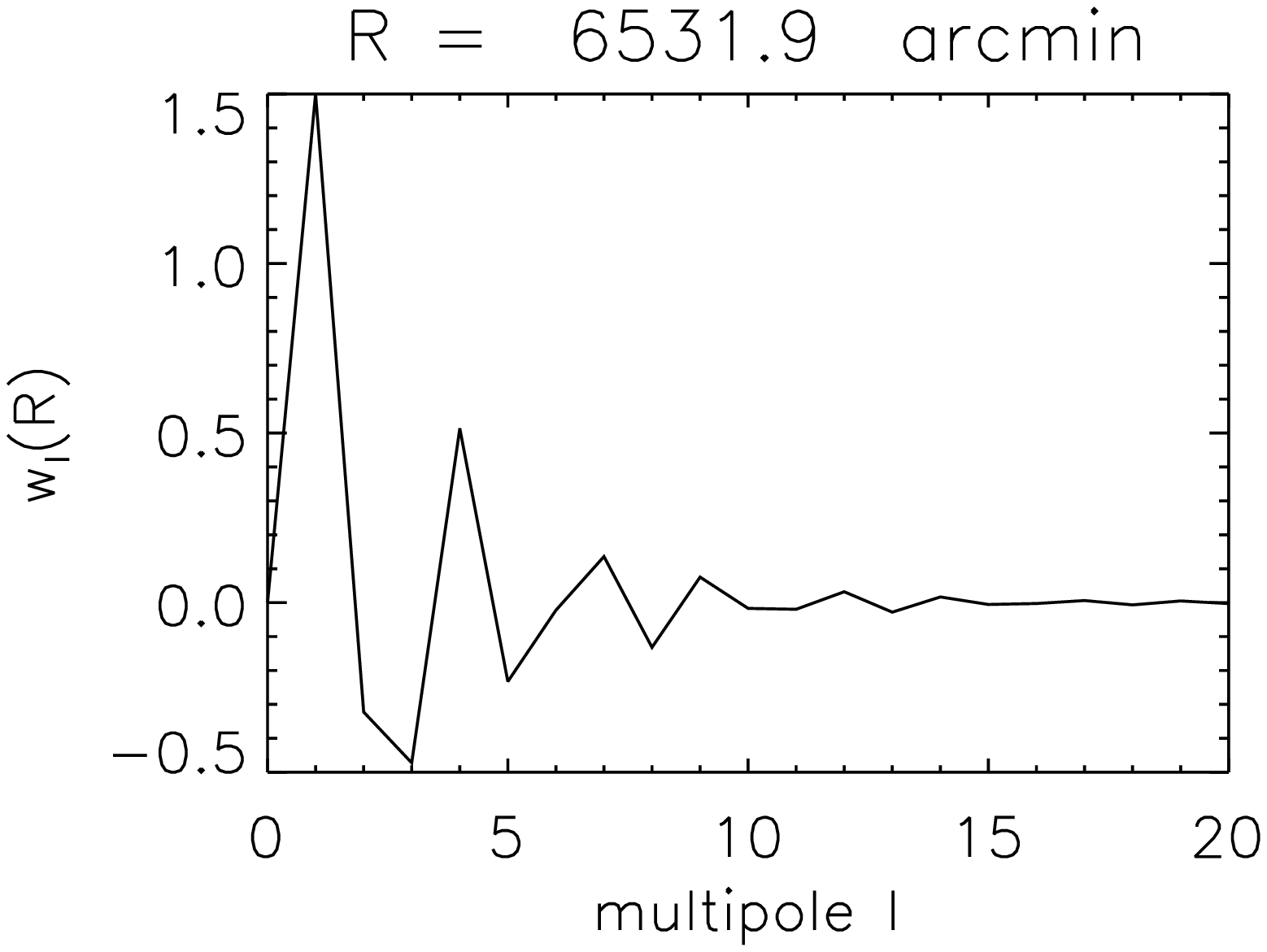}
\includegraphics[width=3.2cm] {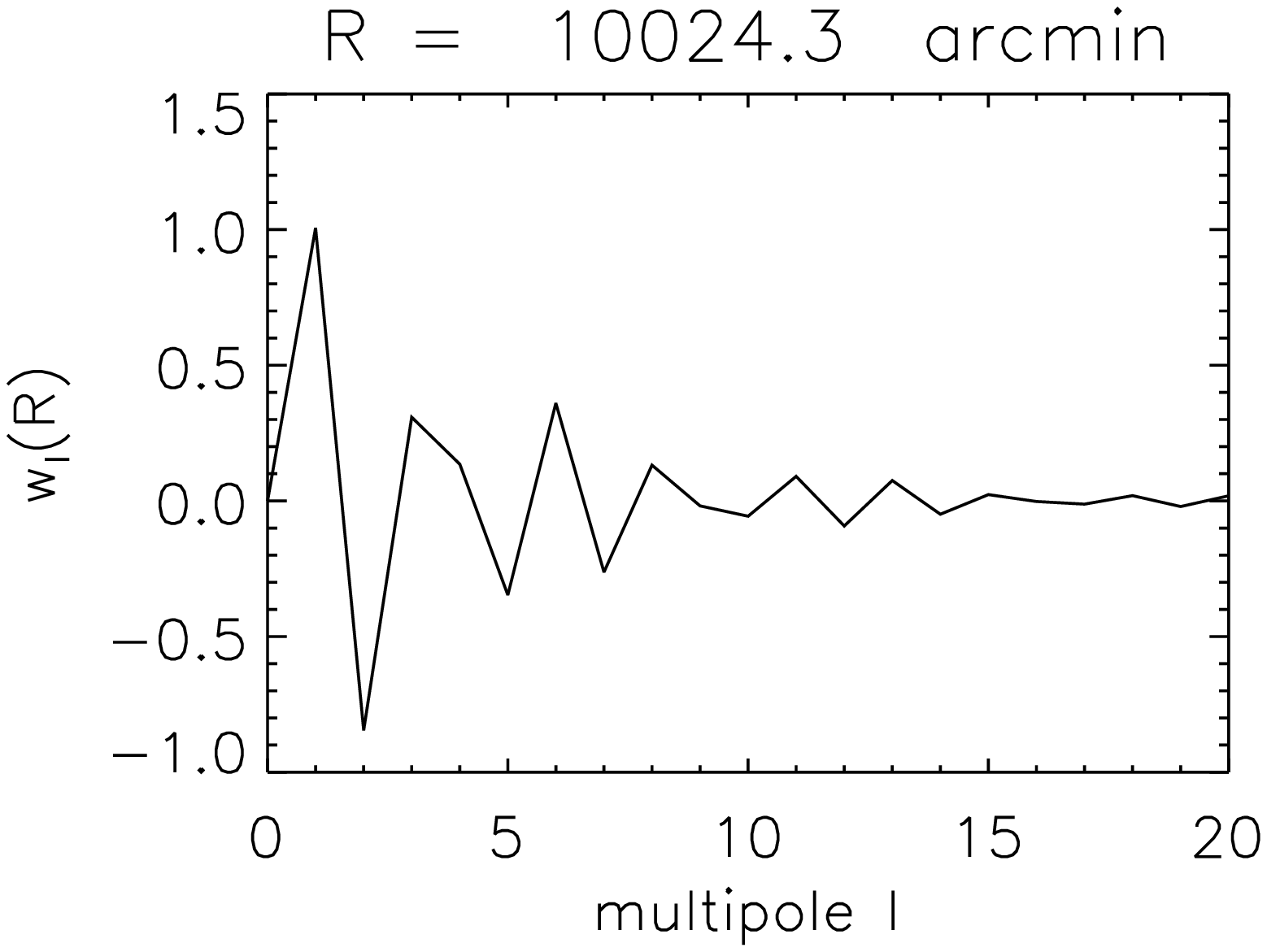}
\caption{The window function of the SMHW for the list of selected
  angular scales for this case. We have convolved each wavelet window
  function with the pixel window function at $N_{side}=512$. Note that
  the scales of 2.9 and 4.5 arc minutes are smaller than the pixel
  size (6.9 arc minutes). However they provide additional information
  because of the shape of the
  wavelet. \label{angular_scales_selected}}
\end{center}
\end{figure*}

Note that as we include higher multipoles (i.e. smaller scales)
$\sigma(f_{nl})$ decreases obtaining a limit of $\sigma(f_{nl}) = 5.4$
for $\ell_{max} = 1535$.

For the wavelet estimator, we consider the ideal experiment and the
cosmological parameters defined previously for the primordial
bispectrum estimators. To maximise the non-Gaussian signal we take a
wide interval of different angular scales. We select 21 angular scales
(including the unconvolved map) from 2.9 arc minutes to 167.07 degrees
logarithmically spaced\footnote{The two smallest scales peak at very
  small angular scales, corresponding to sub-pixel structures at the
  selected resolution of $N_{side}=512$. Although the localization
  properties of the wavelet may be affected for those scales for the
  considered resolution, we have checked that this effect does not
  affect the $f_{nl}$ estimation. To test this point, we have compared
  the values of $\sigma(f_{nl})$ using the same set of 21 angular
  scales in the current case of pixel resultion $N_{side}=512$ and
  $\ell_{max}=1535$ and in another case with $N_{side}=1024$ and
  $\ell_{max}=1535$. The differences in $\sigma(f_{nl})$ are $< 1\%$
  between both cases.}. See Fig. \ref{angular_scales_selected} for the
list of considered scales and the wavelet window function
corresponding to these scales.
%
%
%
%
We compute the third order quantities $\alpha_{ijk}$ given by Eq.
\ref{thealpha_stats} and the covariance matrix of the $q_{ijk}$
statistics given by Eq. \ref{the_cov_qijk_stat} for these angular
scales. There are $(21+3-1)! / [3!(21-1)!] = 1771$ different third
order statistics for 21 angular scales.
\begin{figure*}
  \center
  \includegraphics[height=5.0cm,width=8.0cm]{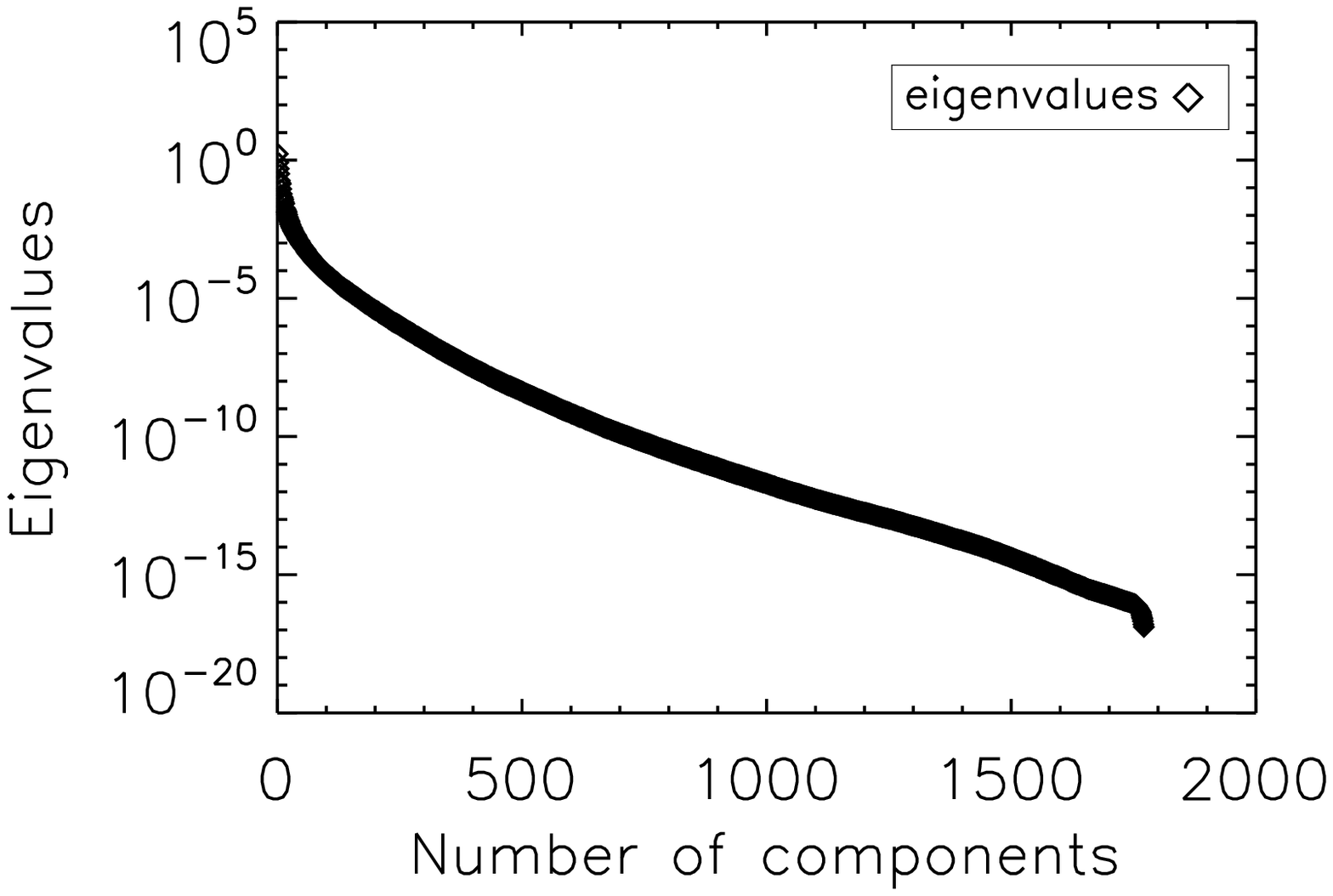}
  \includegraphics[height=5.0cm,width=8.0cm]{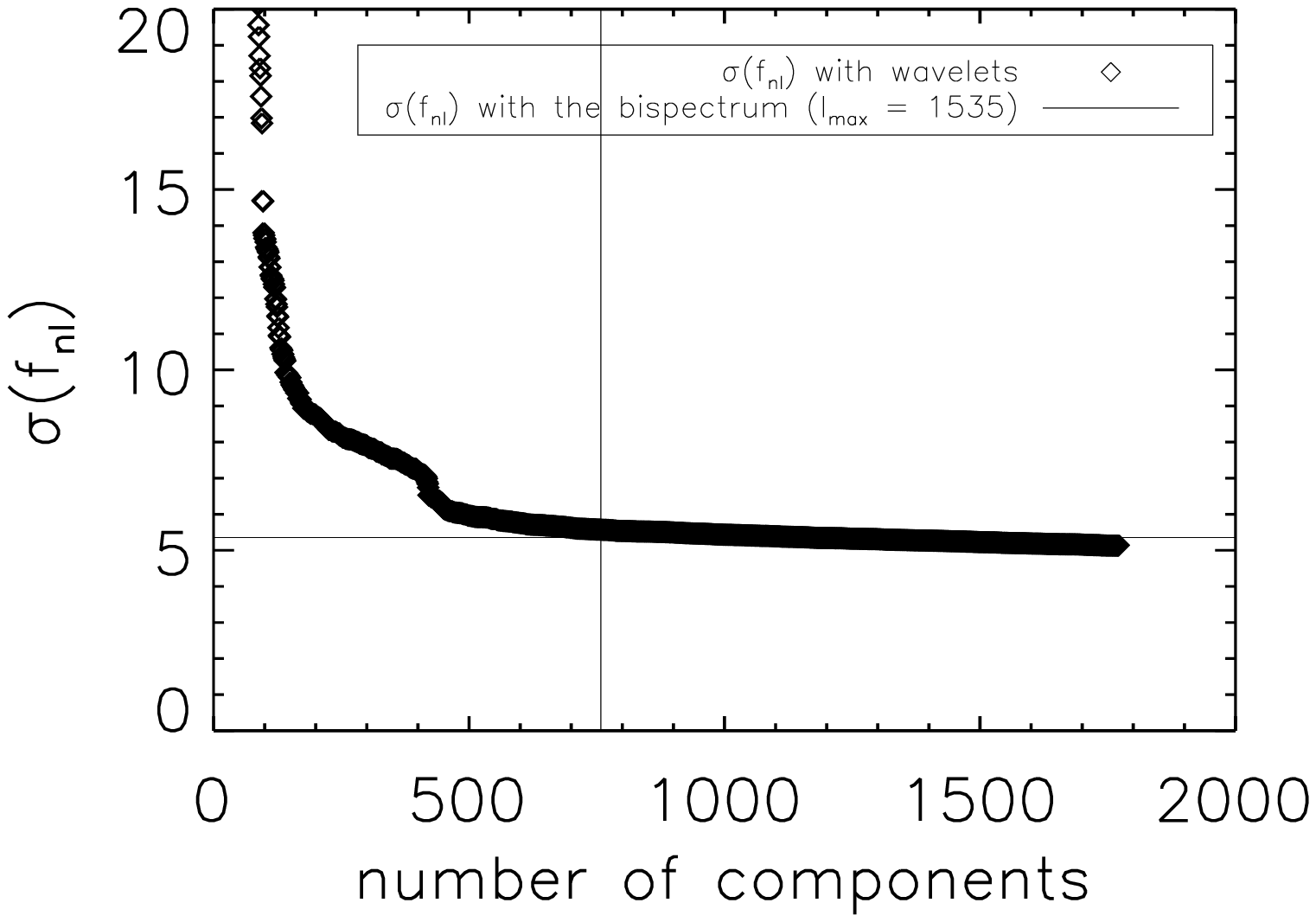}
  \caption{{\it Left:} list of sorted eigenvalues of the covariance
    matrix $C$ among the list of 1771 third order statistics. {\it
      Right:} $\sigma(f_{nl})$ obtained with the wavelet estimator
    using different subsets of eigenvalues of the $C$ matrix following
    the PCA method. The vertical line shows the limit where the ratio
    of the maximum and minimum considered eigenvalues is
    $10^{12}$. \label{eigenvalues_cov_and_sig_fnl}}
\end{figure*}
\begin{figure*}
  \center
  \includegraphics[height=5.0cm,width=8.0cm]{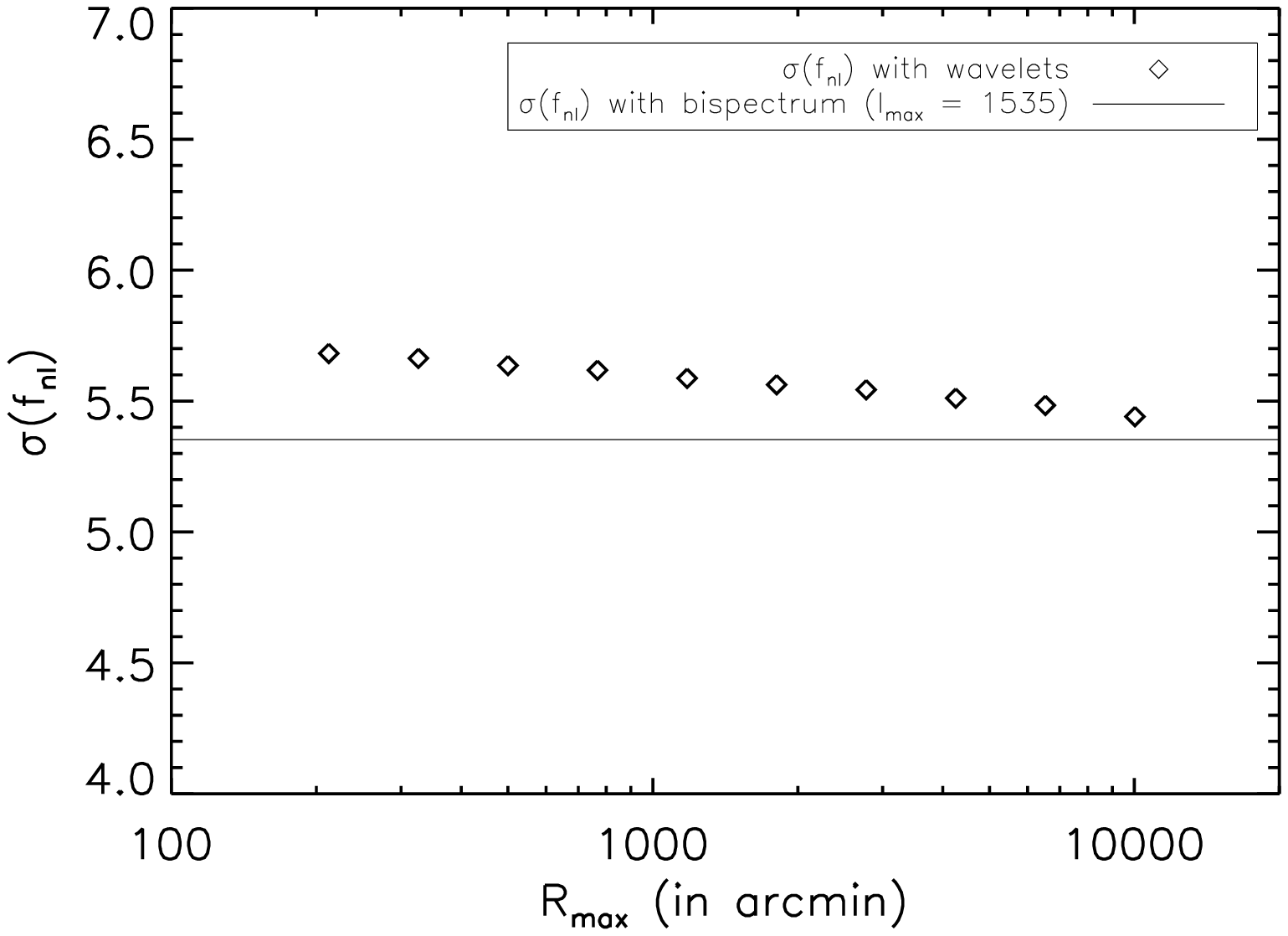}
  \includegraphics[height=5.0cm,width=8.0cm]{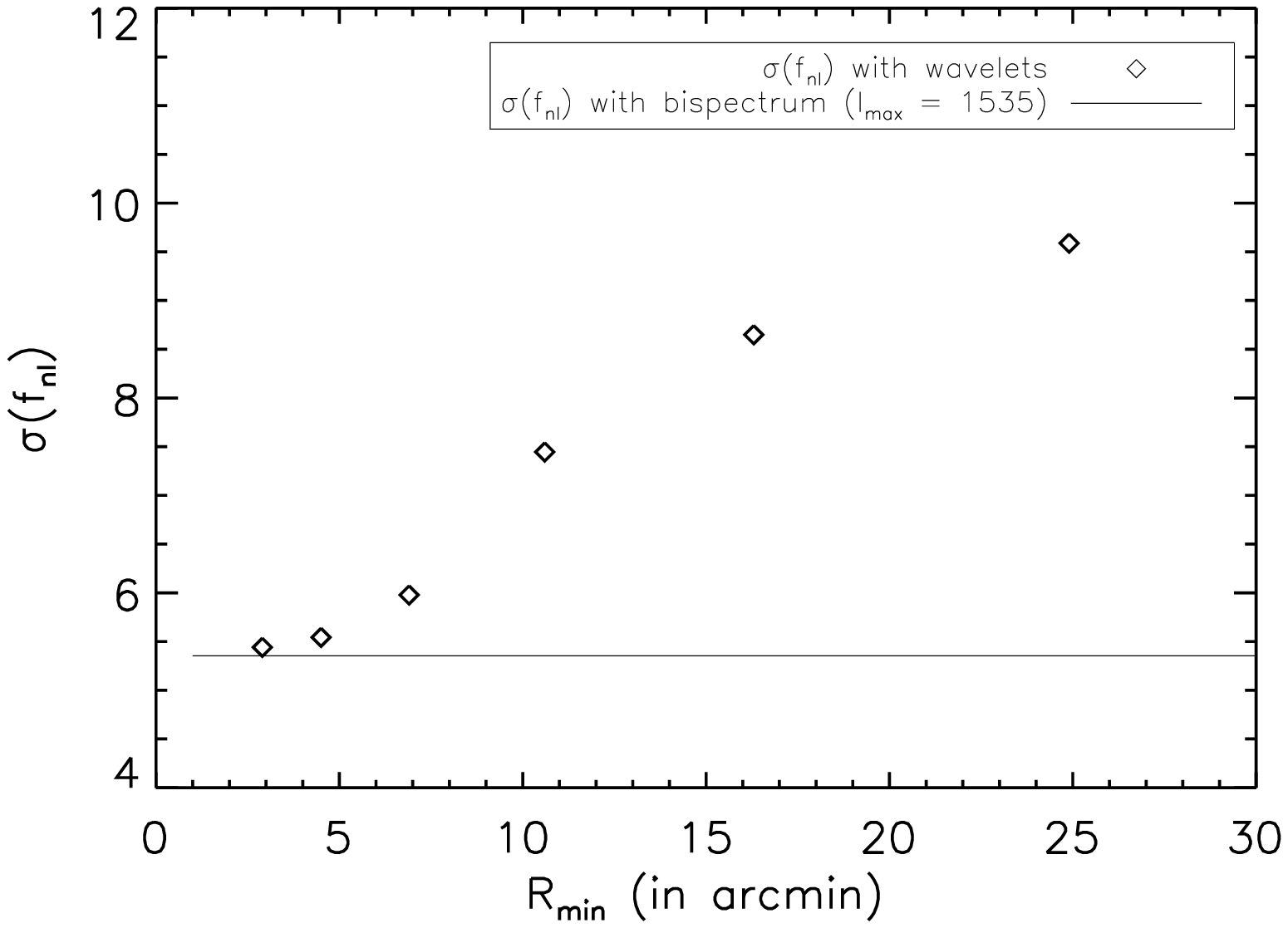}
  \caption{{\it Left:} $\sigma(f_{nl})$ vs the maximum angular scale
    $R_{max}$ considered for the analysis. {\it Right:}
    $\sigma(f_{nl})$ vs the maximum angular scale $R_{min}$ considered
    for the analysis.\label{sig_fnl_vs_scale}}
\end{figure*}
The variance of $f_{nl}$ given by Eq. \ref{sigma_fnl_likelihood} and
the variance estimated with Monte Carlo simulations require the
inverse of the covariance matrix $C_{ijk,rst}$. This covariance matrix
has a large condition number, defined as the ratio of the maximum and
minimum eigenvalues. The eigenvalues of this matrix are plotted on the
left panel of Fig.  \ref{eigenvalues_cov_and_sig_fnl}. This implies
that the computation of its inverse is an ill-conditioned
problem. Therefore the numerical errors present due to the limited
precision of our computers (of the order of $2^{1-53} \simeq
10^{-16}$) can affect the value of $\sigma(f_{nl})$. In order to take
into account the effect of this source of errors, we compute
$\sigma(f_{nl})$ for different subsets of eigenvalues of the $C$
matrix using the PCA described in Subsect. \ref{pca_subsection}. This
is plotted on the right panel of
Fig. \ref{eigenvalues_cov_and_sig_fnl} using the Fisher matrix for
$f_{nl}$ given by Eqs. \ref{sigma_fnl_likelihood_pca1} and
\ref{sigma_fnl_likelihood_pca2} and the PCA method. We as well plot
the value of $\sigma(f_{nl})$ obtained with the Fisher matrix of the
bispectrum given by Eq. \ref{sigma_fnl_bispectrum_intro} for the same
experiment and $\ell_{max}=1535$. In particular, imposing a limit on
the ratio of the minimum and maximum eigenvalues of the covariance
matrix $D_i/D_{max} \le 10^{-12}$, we have $\sigma(f_{nl}) =
$5.4. Similar results are obtained for $\sigma(f_{nl})$ when it is
estimated with $10^3$ Monte Carlo simulations. In particular, we
have obtained $\sigma_{Fisher}(f_{nl})/\sigma_{sims}(f_{nl}) \simeq
0.93$ for the wavelet estimator. We have obtained a similar ratio for
the bispectrum estimator in the same ideal experiment using the Fisher
matrix and Gaussian simulations.

Finally, we have studied the dependence of the wavelet estimator on
the angular scale. In Fig. \ref{sig_fnl_vs_scale} we present the
$\sigma_{Fisher}(f_{nl})$ obtained with the wavelet estimator vs the
$R_{min}$ and the $R_{max}$. We have used the PCA method in order to
minimize the influence of the errors in the inverse of the $C$ matrix
by using only the eigenvalues with $D_i / D_{max} \le 10^{-12}$. We
can see that $\sigma_{Fisher}(f_{nl})$ is highly dependent on
$R_{min}$. This dependence is explained because the small scales map
better the higher multipoles (even for scales smaller than the pixel
size), while the large scales are all centered around low
multipoles. We have also found that the most significant contribution
to the local non-Gaussianity is given by combinations of small and
large scales, which is in agreement with the squeezed configurations
of the shape form of this kind of non-Gaussianity \citep[see for
  example][]{fergusson2009,jeong2009}.
\subsection{Estimated error bars for $f_{nl}$ with a WMAP-like experiment}
We have estimated the expected dispersion of the local $f_{nl}$
parameter for the WMAP 5-year data and WMAP 7-year data. In this
analysis we have considered the instrumental properties of the
experiment, the noise level and the recommended sky cuts of each data
release. We have compared our results with the values obtained with
the optimal estimator for WMAP 5-yr \citep{smith2009}, where
$\sigma(f_{nl}) = 21$, and with the optimal estimator for WMAP 7-yr
\citep{komatsu2010} where $\sigma(f_{nl}) = 21$.
\begin{table*}
  \center
  \caption{Considered angular scales and the fraction of the sky that
    is masked out by each extended mask for the WMAP 5-yr initial
    mask. Similar values are obtained for the WMAP 7-yr
    masks. \label{table_angular_scales_selected_wmap5yr}}
  \begin{tabular}{c|cccccccc}
    \hline \hline
    index $i$ & 0 & 1 & 2 & 3 & 4 & 5 & 6 & 7 \\
    
    scale $R_{i}$ & map & 2.9' & 4.5' & 6.9' & 10.6' & 16.3' & 24.9' & 38.3' \\
    
    masked out area (\%) & 28.4 & 28.4 & 28.4 & 28.8 & 29.3 & 29.9 & 30.7 & 31.8 \\
    \hline
    index $i$ & 8 & 9 & 10 & 11 & 12 & 13 & 14 & - \\
    
    scale $R_{i}$ & 58.7' & 90.1' & 138.3' & 212.3' & 325.8' & 500' & 767.3' & -\\
    
    masked out area (\%) & 33.4 & 36.0 & 40.1 & 46.7 & 55.8 & 68.6 & 85.4 & - \\
    \hline
    \hline 
  \end{tabular}
\end{table*}
\begin{figure*}
  \center
  \includegraphics[height=5.0cm,width=8.0cm]{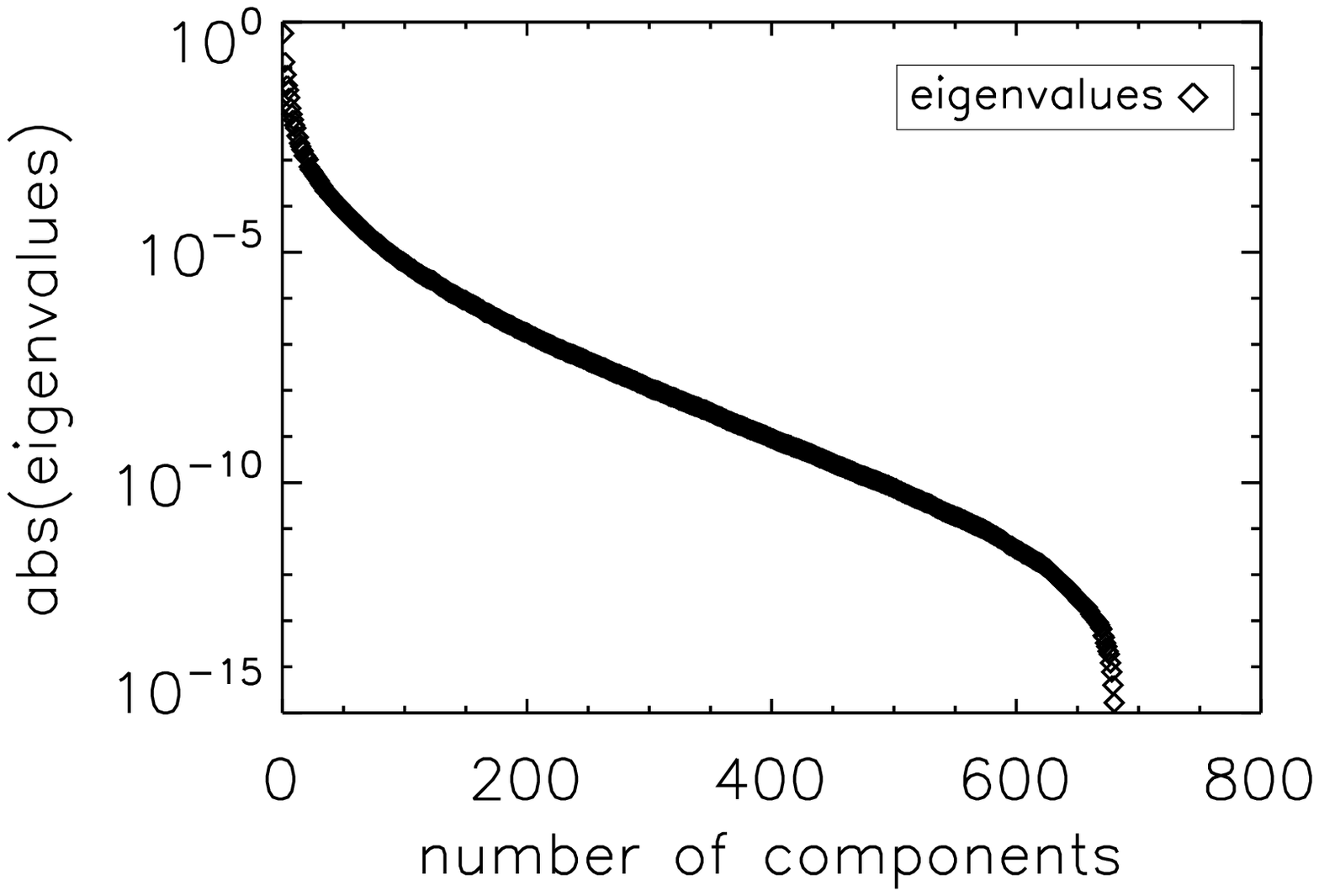}
  \includegraphics[height=5.0cm,width=8.0cm]{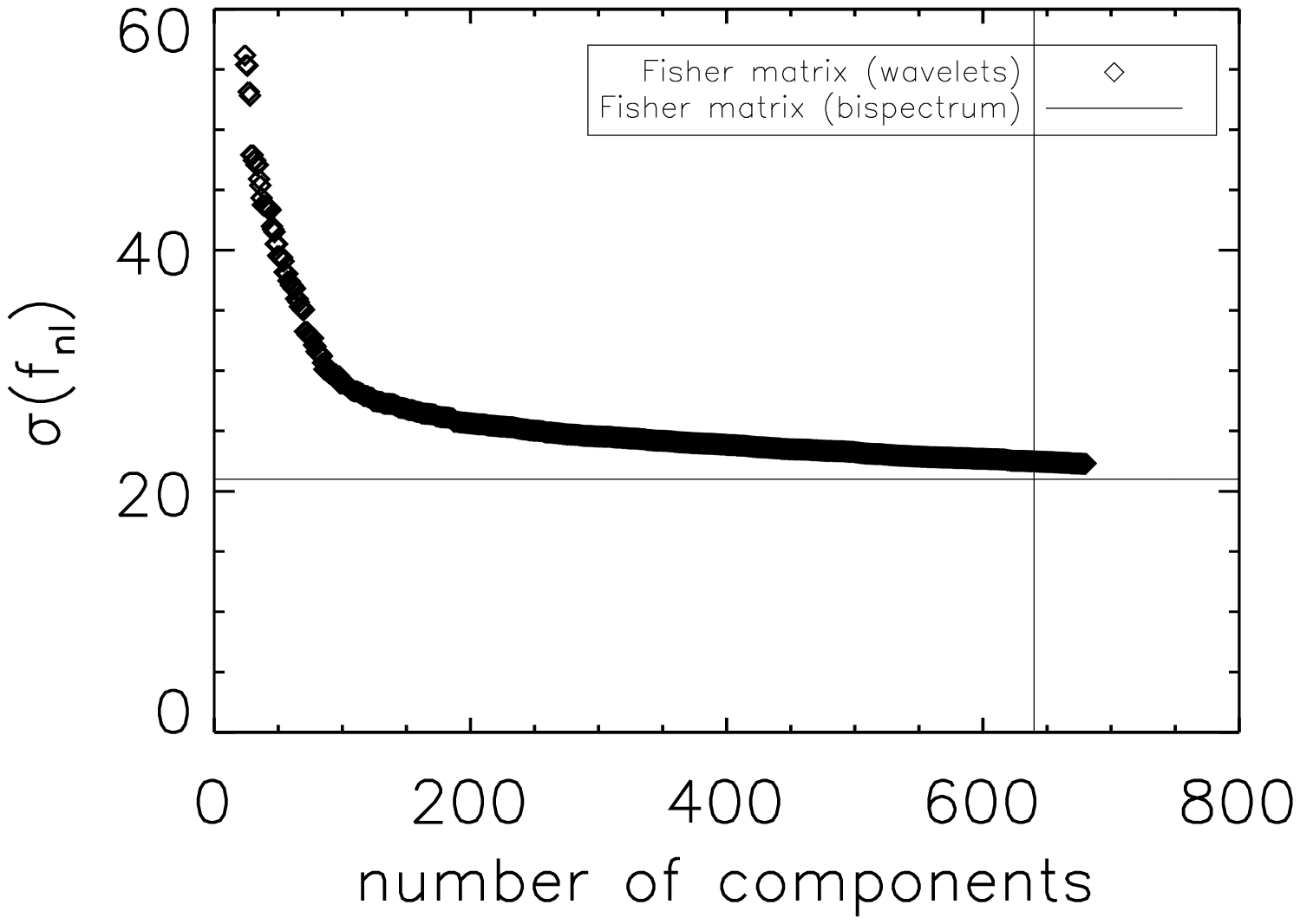}
  \caption{{\it Left:} list of sorted eigenvalues of the covariance
    matrix $C$ among the list of 680 third order statistics (15
    scales) for WMAP 5-year data. {\it Right:} $\sigma(f_{nl})$
    obtained with the wavelet estimator using different subsets of
    eigenvalues of the $C$ matrix following the PCA method. The
    vertical line shows the limit where the ratio of the maximum and
    minimum considered eigenvalues is
    $10^{12}$.\label{eigenvalues_cov_and_sig_fnl_wmap5yr}}
\end{figure*}

For WMAP 5-yr analysis we have used $10^4$ Gaussian simulations of the
combined V+W WMAP 5-yr data to estimate the covariance matrix $C$ and
$300$ non-Gaussian simulations to estimate the expected values of the
third order moments. We have normalised these non-Gaussian simulations
to the $\Lambda$CDM model that best fits the WMAP 5-year data. As we
are using a fraction of the full sky, we need to extend the masks in
order to avoid the propagation of the zeros that are masking the
Galaxy to other regions \citep{curto2009a}, specially for large
angular scales. This imposes a limit on the largest scale available
for the analysis. In particular, from the set of scales given in
Fig. \ref{angular_scales_selected}, we use all the scales with at
least 10\% of the sky. Following the criteria to extend the masks
defined in \citet{curto2009a}, we have 15 available scales. Their
corresponding masked area is given in Table
\ref{table_angular_scales_selected_wmap5yr}. In
Fig. \ref{eigenvalues_cov_and_sig_fnl_wmap5yr} we plot the eigenvalues
of the $C$ matrix for this case and the $\sigma(f_{nl})$ for different
subsets of eigenvalues of $C$ using the PCA method. Using $10^{12}$ as
a safe limit for the ratio of the maximum and minimum eigenvalues, we
obtain $\sigma(f_{nl}) = 22$ for this data. Compared with the optimal
estimator based on the bispectrum \citep{smith2009}, we obtain very
similar results. Compared with the results presented by
\citet{curto2009a} and \citet{curto2009b}, we obtain significant
improvements. The better results obtained here are explained because
of the wider ratio of large-to-small angular scales considered
now. This allows to have more third order moments with squeezed
triangles for their corresponding three angular scales. These are the
combinations where the most significant part of the local
non-Gaussianity signal is located \citep{yadav2010}.
\begin{figure*}
  \center
  \includegraphics[height=5.0cm,width=8.0cm]{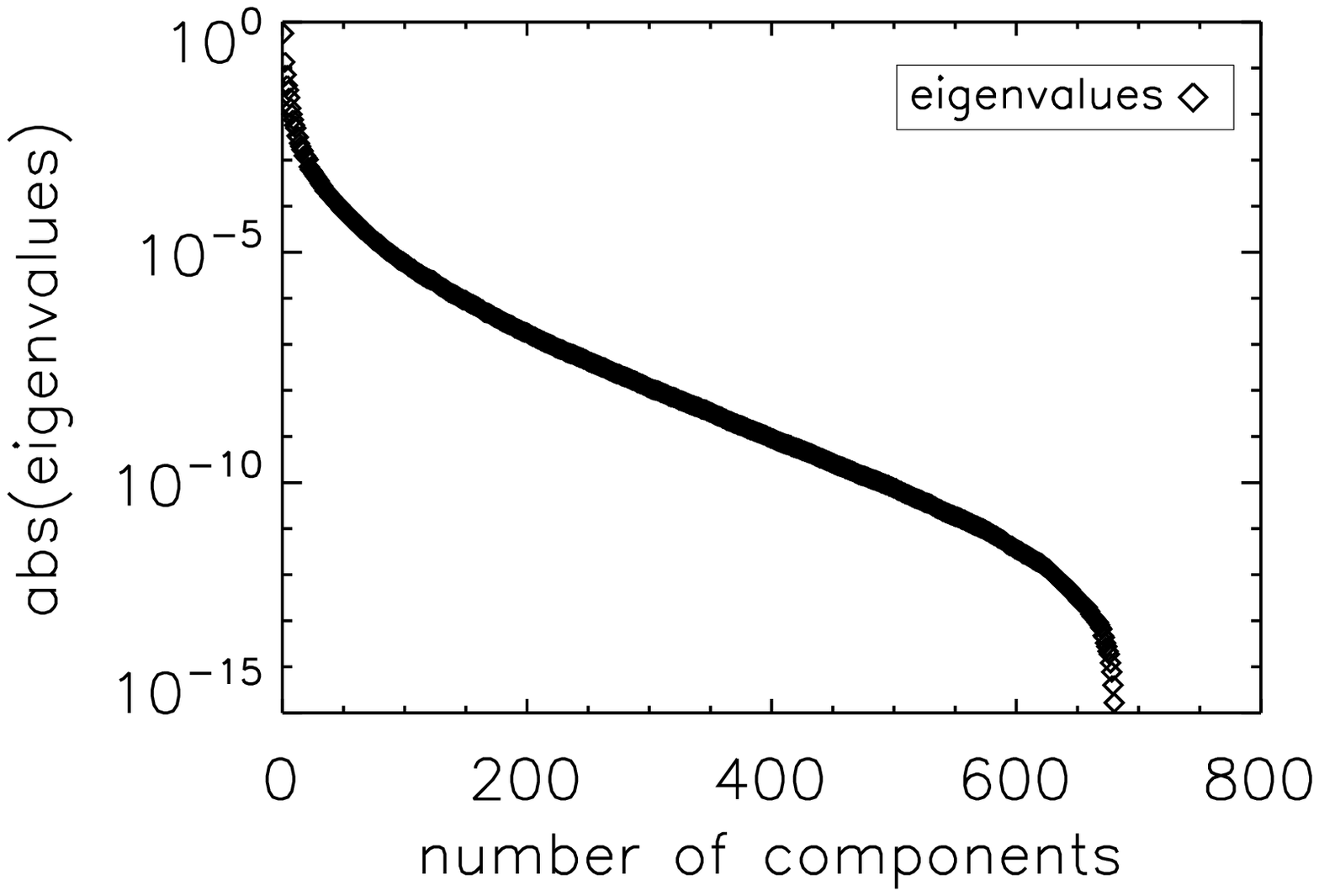}
  \includegraphics[height=5.0cm,width=8.0cm]{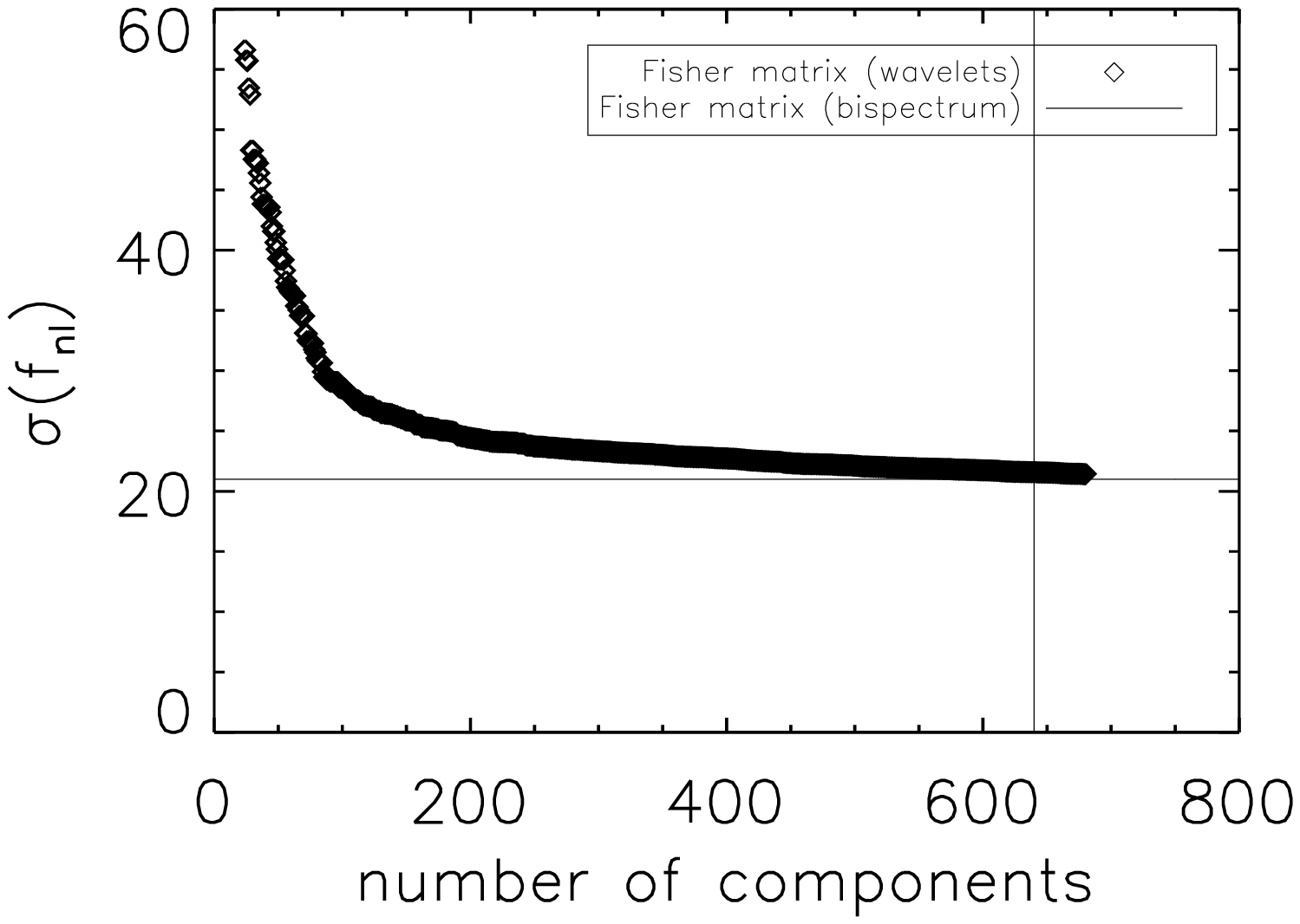}
  \caption{{\it Left:} list of sorted eigenvalues of the covariance
    matrix $C$ among the list of 680 third order statistics (15
    scales) for WMAP 5-year data. {\it Right:} $\sigma(f_{nl})$
    obtained with the wavelet estimator using different subsets of
    eigenvalues of the $C$ matrix following the PCA method. The
    vertical line shows the limit where the ratio of the maximum and
    minimum considered eigenvalues is
    $10^{12}$.\label{eigenvalues_cov_and_sig_fnl_wmap7yr}}
\end{figure*}

For the case of WMAP 7-year data, we have used $10^4$ Gaussian
simulations of the combined V+W WMAP 7-yr data to estimate the
covariance matrix $C$ and the same $300$ non-Gaussian simulations
normalised to the $\Lambda$CDM model that best fits WMAP 7-year data
to estimate the expected values of the third order moments. We
consider the same 15 angular scales as for the WMAP 5-year case
masking a similar area. In
Fig. \ref{eigenvalues_cov_and_sig_fnl_wmap7yr} we plot the eigenvalues
of the $C$ matrix for this case and the $\sigma(f_{nl})$ for different
subsets of eigenvalues of $C$ using the PCA method. Using $10^{12}$ as
a safe limit for the ratio of the maximum and minimum eigenvalues, we
obtain $\sigma(f_{nl}) = 21$ for this data map. Compared with the
optimal estimator based on the bispectrum \citep{komatsu2010}, we
obtain equally optimal error bars with the SMHW.
\section{Conclusions}
\label{conclusions}
We have developed an efficient method to constrain the local
$f_{nl}$ with the CMB anisotropies based on wavelets. We have found
the dependence of the third order moments defined in Eq.
\ref{themoments_qijk} on $f_{nl}$ and the cosmological model through
the primordial bispectrum (see Eq. \ref{thealpha_stats}). On the other
hand we have found an analytical expression for the covariance matrix
for all the third order statistics (see
Eq. \ref{the_cov_qijk_stat}). Assuming a Gaussian-like distribution
for the third order moments, we have estimated the variance of
$f_{nl}$ through the Fisher matrix (see
Subsect. \ref{fisher_wavelets}). This variance is compared with the
variance obtained with the same method for the bispectrum in
Sect. \ref{comparison_with_opt_bisp_est}. Both cases have been applied
to an ideal experiment with an angular resolution of 6.9 arcmin and
without instrumental noise. After applying Principal Component
Analysis in order to minimize the influence of the errors of the
inversion of the covariance matrix, we have found that $\sigma(f_{nl})
= $ 5.4 when we use the wavelets while $\sigma(f_{nl}) = $ 5.4 for the
bispectrum up to $\ell_{max}=1535$. In addition, considering the case
of a more realistic experiment with anisotropic and incomplete sky,
such as the WMAP data, we have obtained $\sigma(f_{nl}) = $ 22 for V+W
WMAP 5-year data and $\sigma(f_{nl}) = $ 21 for V+W WMAP 7-year
data. These results are almost equal to the values obtained with the
optimal bispectrum estimator where $\sigma(f_{nl}) = $ 21
\citep{smith2009,komatsu2010}.  All these results indicate that
wavelets can be as efficient as the bispectrum to detect the
non-Gaussianity of the local type. Apart from the efficiency of the
tool (about 7 times faster than the bispectrum estimator for a
WMAP-like experiment) it is remarkable that, as they are different
statistical estimators, wavelets may be sensitive to different
systematics in real data. Moreover wavelets allow us to test the
isotropic character of the $f_{nl}$ parameter
\citep{curto2009b,rudjord2010} by studying different regions of the
sky. We stress the importance of this statistical tool as an efficient
alternative to measure local $f_{nl}$ in experiments such as
Planck. In forthcoming works we will apply this tool on WMAP real data
to constrain the local and other configurations of the non-linear
coupling parameter $f_{nl}$.
\section*{acknowledgments}
The authors are thankful to Eiichiro Komatsu for his useful comments
that have helped to produce this paper. The authors also thank
Patricio Vielva, Biuse Casaponsa, Michele Liguori, Frode Hansen, and
Sabino Matarrese for useful comments on different computational and
theoretical issues on the primordial non-Gaussianity. We acknowledge
partial financial support from the Spanish Ministerio de Ciencia e
Innovaci\'on project AYA2007-68058-C03-02. A. C. thanks the Spanish
Ministerio de Ciencia e Innovaci\'on for a pre-doctoral fellowship and
the Universidad de Cantabria for a post-doctoral fellowship.
A. C. thanks the Texas Cosmology Center and the University of Texas at
Austin for their hospitality during a research stay in 2009.  The
authors acknowledge the computer resources, technical expertise and
assistance provided by the Spanish Supercomputing Network (RES) node
at Universidad de Cantabria. We acknowledge the use of Legacy Archive
for Microwave Background Data Analysis (LAMBDA). Support for it is
provided by the NASA Office of Space Science. The HEALPix package was
used throughout the data analysis \citep{healpix}.
%
%

%
%
\end{document}